\documentclass[prd,twocolumn,showpacs,preprintnumbers,superscriptaddress,floatfix,nofootinbib]{revtex4-1}

\usepackage{needspace}

\usepackage{graphicx,booktabs}
\graphicspath{{figures/}}
\usepackage{amssymb,amsmath}
\usepackage{txfonts}
\usepackage{graphicx}
\usepackage{dcolumn}
\usepackage{color}
\usepackage{bm}
\usepackage[colorlinks,
citecolor=blue,
anchorcolor=green,
menucolor=orange,
linkcolor=red,
filecolor=red,
runcolor=pink,
urlcolor=blue,
frenchlinks=red]{hyperref}
\usepackage{multirow}
\usepackage[capitalize]{cleveref}
\crefname{figure}{Fig.}{Figs.}
\Crefname{figure}{Fig.}{Figs.}

\usepackage{slashed}
\usepackage{ulem}
\makeatletter

\newcommand{\Rmnum}[1]{\expandafter\@slowromancap\romannumeral #1@}

\makeatother

\begin{document}
	
	\title{\boldmath Role of excited hyperons in
$K^{*}\Xi$ production via $K^{-}N$ scattering}

	\author{Si-Yu Ma}
	\affiliation{Department of Physics, Lanzhou University of Technology,
		Lanzhou 730050, China}
	
	\author{Xiao-Yun Wang}
	\email{xywang@lut.edu.cn}
	\affiliation{Department of Physics, Lanzhou University of Technology,
		Lanzhou 730050, China}
	\affiliation{Lanzhou Center for Theoretical Physics, Key Laboratory of Theoretical Physics of Gansu Province, Lanzhou University, Lanzhou, Gansu 730000, China}

\begin{abstract}

We investigate the production mechanism of $K^-N\to K^*\Xi$ and the role of high-mass excited hyperons within an effective Lagrangian approach, using a combined fit to cross section and polarization data for $K^-p\to K^{*+}\Xi^-$.
The model includes $s$- and $u$-channel exchanges of the ground-state $\Lambda$ and $\Sigma$ hyperons, together with $s$-channel contributions from $\Lambda(2325)$ and $\Sigma(2230)$.
The pronounced near-threshold enhancement of the total cross section is driven mainly by these excited hyperons, whereas $u$-channel exchange dominates at backward angles at higher energies.
Polarization results show that both the
$K^*$ spin-density matrix elements and the $\Xi$ recoil polarization are sensitive to excited-hyperon contributions near threshold.
Notably, the latter remains sensitive to interference involving excited-hyperon amplitudes even at higher energies.
Moreover, with the same fitted parameters and no additional free parameters, the predicted cross sections for $K^-p\to K^{*0}\Xi^0$ and $K^-n\to K^{*0}\Xi^-$ agree well with available measurements.
Finally, a Dalitz analysis of $K^-p \to K^{*+}\Xi^- \to K\pi\Xi^-$, together with existing $K\pi\Xi^-$ cross section data, indicates the feasibility of reconstructing $K^{*+}\Xi^-$ through this final state.
These results suggest that $K^*\Xi$ production provides a valuable reaction channel for investigating high-mass excited hyperons and their hadronic couplings, and offers theoretical guidance for future experimental measurements.

\end{abstract}

	\maketitle

\section{INTRODUCTION}

Hadron spectroscopy provides an important means of exploring the nonperturbative aspects of quantum chromodynamics (QCD)
\cite{ParticleDataGroup:2026,Bulava:2024HadronSpectroscopy,
Klempt:2009pi,Wang:2026Charmonium}.
Hyperon spectroscopy, as an important branch of hadron spectroscopy, offers a useful platform for studying the spectrum of strange baryons and their couplings to different hadronic channels
\cite{Crede:2024Hyperon}.
Compared with the relatively well-established nonstrange baryon spectrum, however, the strange- and multistrange-baryon sectors remain much less understood.
In particular, experimental information on many hyperon states is still limited, especially in the high-mass region, and their resonance properties, hadronic couplings, and production mechanisms remain insufficiently constrained
\cite{Hyodo:2021StrangeBaryon,Thiel:2022LightBaryon}.

In this context, meson-induced reactions, particularly meson--nucleon scattering, provide an important means of investigating the production mechanisms of strange baryons and their excited states \cite{Klempt:2009pi,wl:2024ext}.
$K^-N$ scattering offers two complementary ways to study strange baryons.
On the one hand, it provides direct access to the two-body production of cascade hyperons $\Xi$ with strangeness $S=-2$, whereas many other production mechanisms involve additional particles in the final state
\cite{Sharov:2011xq}.
On the other hand, the antikaon--nucleon system can couple directly to $S=-1$ hyperon resonances, including $\Lambda^*$ and $\Sigma^*$ states.
Studies of final states such as $K\Xi$ can therefore provide further information on the couplings of these resonances to the corresponding hadronic channels
\cite{Kamano:2014zba}.
Previous studies based on hadronic models, coupled-channel approaches, and Regge-based methods have investigated the $K^-N\to K\Xi$ and $K^-N\to K\Xi(1530)$ reactions, providing valuable information on $\Xi$ production and the possible roles of intermediate $\Lambda^{(*)}$ and $\Sigma^{(*)}$ states
\cite{Sharov:2011xq,Kamano:2014zba,Kim:2023DoubleStrangeness,Luo:2026KXi,Guo:2025Xi1530,Wang:2026KXiMolecule}.

In contrast to the more extensively studied $K\Xi$ and $K\Xi(1530)$ channels, $K^*\Xi$ production has received relatively little attention.
Since the $K^*$ is a spin-1 vector meson, its production can provide information on the spin structure of the reaction, vector-meson couplings, and the couplings of excited hyperons to vector-meson final states
\cite{Zhao:2001Kstar,Oh:2006KstarLambda,Kim:2021HyperonPhoto}.
Studies of related $K^*$-associated reactions, including $K^*\Lambda$, $K^*\Sigma$, and $K^*\Sigma^*$, have shown that such final states can provide useful information on strange-hadron production mechanisms and baryon-resonance contributions
\cite{Kim:2011KstarLambda,Wang:2015zcp,Oh:2006Kstar,Xiang:2020phx}.
In addition to the usual cross section observables, the $K^*$ spin-density matrix elements characterize the different helicity components and their coherence, whereas the recoil polarization of the final $\Xi$ reflects the interference among different spin amplitudes.
Previous studies of meson-induced and photoproduction reactions have shown that polarization observables provide important complementary information for analyzing different exchange mechanisms and resonance contributions \cite{Luo:2026KXi,Chen:2020Kstar,Jackson:2015KXi,Anisovich:2017Kstar,Wang:2026KstarSigmaSDME}.
At the same time, the relatively high $K^*\Xi$ production threshold places the reaction in an energy region where high-mass $\Lambda^*$ and $\Sigma^*$ states may contribute.
These features make $K^-N\to K^*\Xi$ not only a means of studying $\Xi$ production but also a complementary channel for exploring high-mass hyperon excitations.

Experimentally, early $K^-N$ scattering measurements provided total cross sections for several $K^*\Xi$ charge channels, together with some differential cross section data
\cite{Badier:1966KstarXi,Haque:1966KstarXi,Abrams:1968KstarXi,Goldwasser:1970Xi,Badier:1972Xi,Mazzucato:1981Xi,Bellefon:1975Xi,London:1966KstarXi,Dauber:1969Xi,Scheuer:1971Kn,DiBianca:1975XiOmega}.
For the $K^-p\to K^{*+}\Xi^-$ reaction, measurements of the $K^*$ spin-density matrix elements and the recoil polarization of the final $\Xi$ are also available \cite{Mazzucato:1981Xi}.
These data provide an important experimental basis for studying the $K^*\Xi$ production mechanism.
The existing measurements, however, remain limited in energy and angular coverage
\cite{Mazzucato:1981Xi,Scheuer:1971Kn},
and the available information is uneven across the different charge channels
\cite{London:1966KstarXi,Dauber:1969Xi,Scheuer:1971Kn,
DiBianca:1975XiOmega}.
A systematic analysis of the available cross section and polarization data is therefore useful for further clarifying the $K^-N\to K^*\Xi$ reaction mechanism and for guiding future experimental studies.

In this work, we investigate the $K^-N\to K^*\Xi$ reactions within an effective Lagrangian approach, focusing on their production mechanisms and the role of high-mass excited hyperons.
The free parameters are determined from a combined fit to the available cross section and polarization data for $K^-p\to K^{*+}\Xi^-$ and used to predict observables for $K^-p\to K^{*0}\Xi^0$ and $K^-n\to K^{*0}\Xi^-$.
We also examine the Dalitz process $K^-p\to K^{*+}\Xi^-\to K\pi\Xi^-$.
These analyses allow us to test the applicability of the model across different charge channels, assess the sensitivity of polarization observables to excited-hyperon contributions, and explore the feasibility of reconstructing $K^{*+}\Xi^-$ in the $K\pi\Xi^-$ final state.

The remainder of this paper is organized as follows.
\Cref{zhangjie2} presents the theoretical framework
and observables, \Cref{zhangjie3} discusses the fit results and predictions for the related reactions, \Cref{zhangjie4}  examines the Dalitz process,
and \cref{zhangjie5} summarizes the findings.

	\section{PRODUCTION OF $K^*\Xi$ VIA $K^-N$ SCATTERING}
\label{zhangjie2}
\begin{figure}[hbp]
	\centering
	\includegraphics[width=0.95\linewidth]{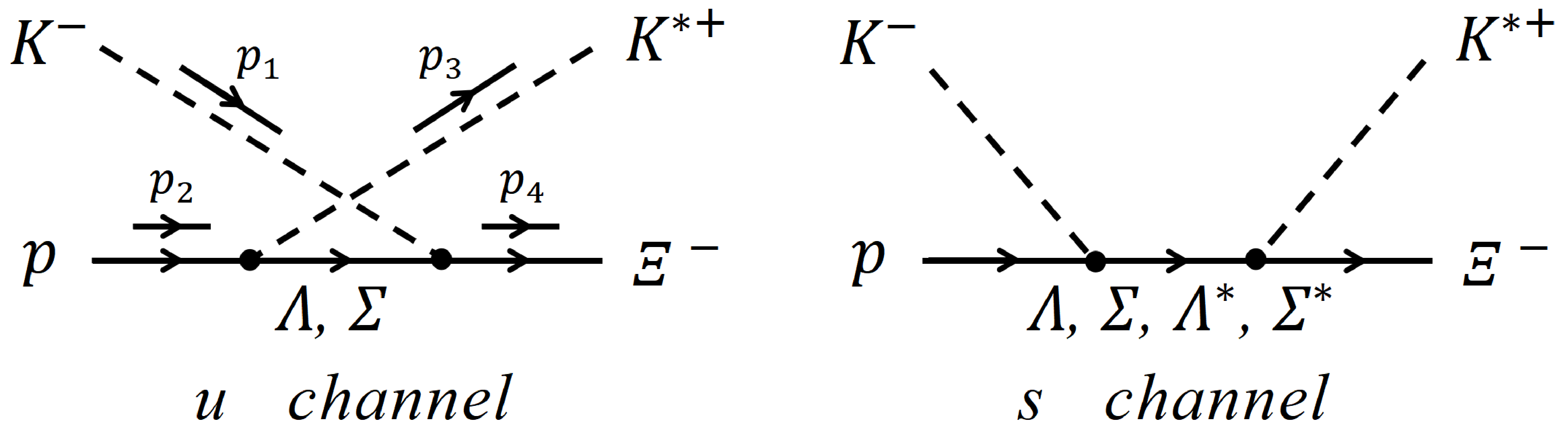}
	\caption{Feynman diagrams for the
	$K^-p\to K^{*+}\Xi^-$ reaction.}
	\label{fig1}
\end{figure}

In this work, we employ an effective Lagrangian approach to study the $K^-N\to K^*\Xi$ reactions, with the $K^-p\to K^{*+}\Xi^-$ channel serving as the primary reaction for detailed analysis.
The corresponding tree-level Feynman diagrams are shown in \cref{fig1}.

Specifically, we include the ground-state $\Lambda$ and $\Sigma$ hyperon exchanges in both the $s$- and $u$-channels, together with the $s$-channel exchanges of the excited hyperons $\Lambda(2325)$
and $\Sigma(2230)$.
The available total cross section data for $K^-p\to K^{*+}\Xi^-$ show a rapid rise above threshold and a pronounced peak in the near-threshold region \cite{Bellefon:1975Xi,Dauber:1969Xi,London:1966KstarXi}.
The masses of these two excited hyperons lie close to the $K^*\Xi$ production threshold.
They have isospins $I=0$ and $I=1$, respectively, and their lowest allowed partial waves in the $K^*\Xi$ final state are $S$ and $P$ waves, respectively \cite{ParticleDataGroup:2026}.
We therefore take $\Lambda(2325)$ and $\Sigma(2230)$ as representative candidates for the $s$-channel contributions.
Given the limited experimental data, we do not introduce additional excited hyperons to keep the number of free parameters manageable.
Although these excited hyperons can, in principle, also be exchanged in the $u$-channel, their larger masses place the corresponding poles far from the physical $u$-channel region, leading to strong suppression of the corresponding exchange amplitudes. Therefore, their $u$-channel contributions are neglected in the present work \cite{Luo:2026KXi}.
In addition, since a conventional $q\bar q$ meson cannot carry $|S|=2$, no conventional one-meson-exchange contribution occurs in the $t$-channel~\cite{Sharov:2011xq}.
For convenience, the ground-state $\Lambda$ and $\Sigma$ hyperons are collectively denoted by $Y$, while $\Lambda(2325)$ and $\Sigma(2230)$ are denoted by $Y^*$.
The intermediate hyperons included in the calculation and their basic properties are summarized in \cref{tab1}.

\begin{table}[tbp]
	\renewcommand{\arraystretch}{1.5}
	\caption{Intermediate hyperons included in the present calculation.
	The listed properties are taken from the PDG
	\cite{ParticleDataGroup:2026}.}

    \label{tab1}
	\centering
	\footnotesize
	\setlength{\tabcolsep}{7pt}
	\begin{tabular}{ccccc}
		\hline\hline
		State
		& $J^P$
		& $M$ (MeV)
		& $\Gamma$ (MeV)
		& PDG Rating \\
		\hline
		$\Lambda(1116)$
		& $\frac{1}{2}^{+}$
		& 1116
		& --
		& $\star\star\star\star$ \\
        \hline
		$\Sigma(1193)$
		& $\frac{1}{2}^{+}$
		& 1193
		& --
		& $\star\star\star\star$ \\
        \hline
		$\Lambda(2325)$
		& $\frac{3}{2}^{-}$
		& 2325
		& 177
		& $\star$ \\
        \hline
		$\Sigma(2230)$
		& $\frac{3}{2}^{+}$
		& 2240
		& 350
		& $\star$ \\
		\hline\hline
	\end{tabular}
\end{table}

To construct the amplitudes corresponding to \cref{fig1}, we adopt effective Lagrangian forms commonly used in related studies
\cite{Xiang:2020phx,Nakayama:2006tyg,Gao:2010hy,Wang:2019dsi,Wang:2024KstarSigma,Wang:2024BESIII,Kim:2024HighSpin,Pan:2025Lambda}.
Here, $Y_{1/2^+}$ denotes a ground-state $\Lambda$ or $\Sigma$ hyperon, while $Y^*_{3/2^\pm}$ denotes a spin-$3/2$ excited hyperon with positive or negative parity.
The effective Lagrangians relevant to the $u$-channel contributions are written as
\begin{eqnarray}
	\mathcal{L}_{K\Xi Y_{1/2^+}}
	&=&
	-i g_{K\Xi Y_{1/2^+}}
	\bar{\Xi}\gamma_5Y_{1/2^+}K
	+\mathrm{H.c.},
	\label{2 lag:KXiY}	
	\\
\mathcal{L}_{NK^*Y_{1/2^+}}
    &=&
	-g_{NK^*Y_{1/2^+}}
    \bar{N}\left(
    \slashed{K}^{*}-
    \frac{\kappa_{NK^*Y_{1/2^+}}}{m_N}
	\sigma_{\mu\nu}\partial^\nu K^{*\mu}
	\right)Y_{1/2^+}
	\notag\\&&+\mathrm{H.c.}
	\label{1 lag:NKstarY}   
\end{eqnarray}
For the $s$-channel, the effective Lagrangians read
\begin{eqnarray}
	\mathcal{L}_{K^*\Xi Y_{1/2^+}}
	&=&
	-g_{K^*\Xi Y_{1/2^+}}
	\bar{\Xi}
	\left(
	\slashed{K}^{*}
	-\frac{\kappa_{K^*\Xi Y_{1/2^+}}}{m_N}
	\sigma_{\mu\nu}\partial^\nu K^{*\mu}
	\right)
	Y_{1/2^+}
	\notag\\
	&&
	+\mathrm{H.c.},
	\label{4 lag:KstarXiY}
	\\	
  	\mathcal{L}_{KNY_{1/2^+}}
	&=&
	-i g_{KNY_{1/2^+}}
	\bar{N}\gamma_5Y_{1/2^+}K
	+\mathrm{H.c.},
	\label{3 lag:KNY}
    \\
	 \mathcal{L}_{K^*\Xi Y^*_{3/2^\pm}}
	&=&
	-i\frac{g^1_{K^*\Xi Y^*_{3/2^\pm}}}{2m_\Xi}
	\bar{\Xi}\gamma^\mu
	\left\{
	\begin{array}{c}
		\gamma_5 \\
		1
	\end{array}
	\right\}
	K^*_{\mu\nu}Y^{*\nu}_{3/2^\pm}
	\notag\\
	&&
	-\frac{g^2_{K^*\Xi Y^*_{3/2^\pm}}}{(2m_\Xi)^2}
	(\partial^\nu\bar{\Xi})
	\left\{
	\begin{array}{c}
		\gamma_5 \\
		1
	\end{array}
	\right\}
	K^*_{\mu\nu}Y^{*\mu}_{3/2^\pm}
	\notag\\
	&&
	+\frac{g^3_{K^*\Xi Y^*_{3/2^\pm}}}{(2m_\Xi)^2}
	\bar{\Xi}
	\left\{
	\begin{array}{c}
		\gamma_5 \\
		1
	\end{array}
	\right\}
	(\partial^\nu K^*_{\mu\nu})
	Y^{*\mu}_{3/2^\pm}
	+\mathrm{H.c.},
	\label{6 lag:KstarXiYstar}
	\\
	\mathcal{L}_{KNY^*_{3/2^\pm}}
	&=&
	\frac{g_{KNY^*_{3/2^\pm}}}{m_K}
	\bar{N}
	\left\{
	\begin{array}{c}
		1 \\
		\gamma_5
	\end{array}
	\right\}
	Y^{*\mu}_{3/2^\pm}\partial_\mu K
	+\mathrm{H.c.},
	\label{5 lag:KNYstar}   
\end{eqnarray}
where $K^*_{\mu\nu}=\partial_\mu K^*_\nu-\partial_\nu K^*_\mu$. In Eqs.~\eqref{6 lag:KstarXiYstar} and \eqref{5 lag:KNYstar}, the upper and lower entries correspond to positive- and negative-parity $Y^*$ resonances, respectively.
For isovector $\Sigma$ exchange, the corresponding isospin factors are introduced through the Pauli matrices $\tau$.
The fields $K$ and $K^*$ denote the pseudoscalar $K$ meson and the $K^*(892)$ vector meson,  while $N$, $\Xi$, $Y$, and $Y^*$ denote the nucleon, cascade-hyperon, ground-state-hyperon, and excited-hyperon fields, respectively.
For the $K^*\Xi Y^*$ vertex, only the leading coupling $g^1_{K^*\Xi Y^*}$ is retained.
The higher-derivative couplings $g^2_{K^*\Xi Y^*}$ and $g^3_{K^*\Xi Y^*}$ are set to zero because of the lack of experimental constraints
\cite{Xiang:2020phx,Kim:2024HighSpin}.

The coupling constants associated with the ground-state hyperon vertices are taken from previous studies
\cite{Nakayama:2006tyg,Wang:2024BESIII}.
For the $NK^*Y$ vertices, we use
$g_{NK^*\Lambda}=-6.11$,
$\kappa_{NK^*\Lambda}=2.43$,
$g_{NK^*\Sigma}=-3.52$, and
$\kappa_{NK^*\Sigma}=-1.14$.
For the $K\Xi Y$ and $KNY$ vertices, we take
$g_{K\Xi\Lambda}=3.52$,
$g_{K\Xi\Sigma}=-13.26$,
$g_{KN\Lambda}=-13.24$, and
$g_{KN\Sigma}=3.58$.
For the $K^*\Xi Y$ vertices, we use
$g_{K^*\Xi\Lambda}=6.11$,
$\kappa_{K^*\Xi\Lambda}=0.65$,
$g_{K^*\Xi\Sigma}=-3.52$, and
$\kappa_{K^*\Xi\Sigma}=4.22$.
For the $KNY^*$ vertices, the magnitudes of the coupling constants are determined from the partial decay widths
$Y^*\to\bar{K}N$.
For spin-$3/2$ hyperons, the corresponding decay-width formulas are given by 
\cite{Xiang:2020phx,Kim:2024HighSpin}
\begin{eqnarray}
	\Gamma_{\Lambda(2325)\to\bar{K}N}
	&=&
	\frac{g_{KN\Lambda(2325)}^2}{12\pi}
	\frac{
	\left|\bm{p}_N^{\,\mathrm{c.m.}}\right|^3
	\left(E_N-m_N\right)
	}{
	m_K^2m_{\Lambda(2325)}
	},
	\label{7 width:Lambda2325}
	\\
	\Gamma_{\Sigma(2230)\to\bar{K}N}
	&=&
	\frac{g_{KN\Sigma(2230)}^2}{12\pi}
	\frac{
	\left|\bm{p}_N^{\,\mathrm{c.m.}}\right|^3
	\left(E_N+m_N\right)
	}{
	m_K^2m_{\Sigma(2230)}
	},
	\label{8 width:Sigma2230}
\end{eqnarray}
where
\begin{gather}
\left|\bm{p}_N^{\,\mathrm{c.m.}}\right|
=
\frac{
\sqrt{\lambda\!\left(m_{Y^*}^2,m_N^2,m_K^2\right)}
}{
2m_{Y^*}
},\quad
\;
E_N=
\sqrt{
\left|\bm{p}_N^{\,\mathrm{c.m.}}\right|^2+m_N^2
},
\label{9 momentum:YNK}
\\
\lambda(x,y,z)
=
(x-y-z)^2-4yz.
\label{10 Kallen:function}
\end{gather}
The total widths and the $\bar{K}N$ branching fractions adopted here are taken from the PDG \cite{ParticleDataGroup:2026}:
$\Gamma_{\Lambda(2325)}=177$ MeV, $\mathrm{Br}[\Lambda(2325)\to\bar{K}N]=0.19$, $\Gamma_{\Sigma(2230)}=350$ MeV, and $\mathrm{Br}[\Sigma(2230)\to\bar{K}N]=0.06$.
The corresponding partial decay widths are $\Gamma_{\Lambda(2325)\to\bar{K}N}=33.63$ MeV and $\Gamma_{\Sigma(2230)\to\bar{K}N}=21.00$ MeV, which yield $g_{KN\Lambda(2325)}=1.656$ and $g_{KN\Sigma(2230)}=0.571$, respectively. Since the partial decay widths determine only the magnitudes
of the coupling constants, their signs are taken to be positive by convention in the present calculation \cite{Kim:2024HighSpin,Xie:2014Delta1940}. The remaining couplings
$g^1_{K^*\Xi\Lambda(2325)}$ and
$g^1_{K^*\Xi\Sigma(2230)}$ are treated as free parameters and determined by fitting the experimental data.

To account for the finite size of hadrons, we introduce the following phenomenological form factor for the $s$- and $u$-channel hyperon exchanges \cite{Wang:2015zcp,Gao:2010hy}.
\begin{equation}
	F_i(q_i)
	=
	\frac{\Lambda_i^4}
	{\Lambda_i^4+\left(q_i^2-m_i^2\right)^2},
	\label{11 formfactor}
\end{equation}
where $q_i$ and $m_i$ denote the four-momentum and mass of the exchanged hyperon, respectively. The cutoff parameters $\Lambda_u$, $\Lambda_{s,Y}$, and $\Lambda_{s,Y^*}$ correspond to the $u$-channel ground-state hyperons, the $s$-channel ground-state hyperons, and the $s$-channel excited hyperons, respectively, and are determined by fitting the experimental data.

With the above Lagrangians and form factors, the spin-resolved
amplitude for the $K^-p\to K^{*+}\Xi^-$ reaction is written as
\begin{equation}
\mathcal M_{\lambda s_\Xi s_N}
=
\epsilon_\mu^*(p_3,\lambda)
\bar u(p_4,s_\Xi)
\mathcal A^\mu
u(p_2,s_N).
\label{eq:amplitude}
\end{equation}
Here $\epsilon_\mu^*(p_3,\lambda)$ is the polarization vector of the outgoing
$K^{*+}$ meson with helicity $\lambda=+1,0,-1$, while $s_\Xi$ and $s_N$
denote the spin projections of the final $\Xi$ and the initial nucleon,
respectively. The reduced amplitude $\mathcal A^\mu$ is constructed as the
coherent sum
\begin{align}
\mathcal A^\mu
={}&
\mathcal A^\mu_{u,\Lambda}
+\mathcal A^\mu_{u,\Sigma}
+\mathcal A^\mu_{s,\Lambda}
+\mathcal A^\mu_{s,\Sigma}
\nonumber\\
&+
e^{i\phi_{\Lambda^*}}
\mathcal A^\mu_{s,\Lambda(2325)}
+
e^{i\phi_{\Sigma^*}}
\mathcal A^\mu_{s,\Sigma(2230)},
\label{eq:reduced-amplitude}
\end{align}
where the phase factors $e^{i\phi_{\Lambda^*}}$ and
$e^{i\phi_{\Sigma^*}}$ are introduced phenomenologically to account for
dynamical phase effects that are not explicitly included at the tree level
\cite{Luo:2026KXi}.
The phases $\phi_{\Lambda^*}$ and $\phi_{\Sigma^*}$ are treated as free
parameters in the fit.
No additional adjustable phases are assigned to the ground-state hyperon
exchanges; their fixed amplitude convention defines the reference for these
relative phases.

The reduced amplitudes for the $u$- and $s$-channel contributions
are given by
\begin{eqnarray}
	\mathcal{A}^{\mu}_{u,\Lambda}
	&=& -i g_{K\Xi\Lambda}g_{NK^*\Lambda}F_u(q_u)\gamma^5
	\frac{\slashed{q}_u+m_\Lambda}{u-m_\Lambda^2}
	\notag\\
    &&\times\left[\gamma^\mu+\frac{\kappa_{NK^*\Lambda}}{2m_N}
	\left(\gamma^\mu\slashed{p}_3-\slashed{p}_3\gamma^\mu\right)\right],
	\label{14 amp:uLambda}\\
	\mathcal{A}^{\mu}_{u,\Sigma}
	\!&=&\! -i g_{K\Xi\Sigma}g_{NK^*\Sigma}F_u(q_u)\gamma^5
	\frac{\slashed{q}_u+m_\Sigma}{u-m_\Sigma^2}
	\notag\\
	&&\times\left[\gamma^\mu+\frac{\kappa_{NK^*\Sigma}}{2m_N}
	\left(\gamma^\mu\slashed{p}_3-\slashed{p}_3\gamma^\mu\right)\right],
	\label{15 amp:uSigma}\\
	\mathcal{A}^{\mu}_{s,\Lambda}
	\!&=&\! -i g_{KN\Lambda}g_{K^*\Xi\Lambda}F_{s,Y}(q_s)
	\notag\\
	&&\times\left[\gamma^\mu+\frac{\kappa_{K^*\Xi\Lambda}}{2m_N}
	\left(\gamma^\mu\slashed{p}_3-\slashed{p}_3\gamma^\mu\right)\right]
    \frac{\slashed{q}_s+m_\Lambda}{s-m_\Lambda^2}\gamma^5,
	\label{16 amp:sLambda}\\
    %
	\mathcal{A}^{\mu}_{s,\Sigma}
	\!&=&\! -i g_{KN\Sigma}g_{K^*\Xi\Sigma}F_{s,Y}(q_s)
	\notag\\
	&&\times\left[\gamma^\mu+\frac{\kappa_{K^*\Xi\Sigma}}{2m_N}
	\left(\gamma^\mu\slashed{p}_3-\slashed{p}_3\gamma^\mu\right)\right]
	\frac{\slashed{q}_s+m_\Sigma}{s-m_\Sigma^2}\gamma^5,
	\label{17 amp:sSigma}\\
%
  \mathcal{A}^{\mu}_{s,\Lambda(2325)}
	\!&=&\! i \frac{g^1_{K^*\Xi\Lambda(2325)}}{2m_\Xi}
	\frac{g_{KN\Lambda(2325)}}{m_K}F_{s,Y^*}(q_s)
	\left(p_3^\alpha\gamma^\mu-\slashed{p}_3g^{\mu\alpha}\right)
	\notag\\
	&&\times\frac{\slashed{q}_s+m_{\Lambda(2325)}}
	{s-m_{\Lambda(2325)}^2+i m_{\Lambda(2325)}\Gamma_{\Lambda(2325)}}
	\notag\\
	&&\times\Delta_{\alpha\beta}(q_s)\gamma^5p_1^\beta,
	\label{18 amp:sLambda2325}\\
	\mathcal{A}^{\mu}_{s,\Sigma(2230)}
	\!&=&\! i \frac{g^1_{K^*\Xi\Sigma(2230)}}{2m_\Xi}
	\frac{g_{KN\Sigma(2230)}}{m_K}F_{s,Y^*}(q_s)\gamma^5\!
	\left(p_3^\alpha\gamma^\mu-\slashed{p}_3g^{\mu\alpha}\right)
	\notag\\
	&&\times\frac{\slashed{q}_s+m_{\Sigma(2230)}}
	{s-m_{\Sigma(2230)}^2+i m_{\Sigma(2230)}\Gamma_{\Sigma(2230)}}
	\notag\\
	&&\times\Delta_{\alpha\beta}(q_s)p_1^\beta.
	\label{19 amp:sSigma2230}
\end{eqnarray}
The spin-$3/2$ projection operator is given by
\begin{eqnarray}
\Delta_{\alpha\beta}(q_s)
&=&\!
-g_{\alpha\beta}
+\frac{1}{3}\gamma_\alpha\gamma_\beta
+\frac{1}{3m_{Y^*}}
\left(
\gamma_\alpha q_{s\beta}
-q_{s\alpha}\gamma_\beta
\right)
\notag\\
&&+\frac{2}{3m_{Y^*}^{2}}
q_{s\alpha}q_{s\beta},
\label{20 eq:projector}
\end{eqnarray}
where
$q_s=p_1+p_2$,
$q_u=p_2-p_3=p_4-p_1$,
$s=q_s^2$,
$u=q_u^2$,
and $m_{Y^*}$ denotes the mass of the corresponding excited
hyperon.

\begin{figure}[t]
	\centering
	\includegraphics[width=0.98\columnwidth]{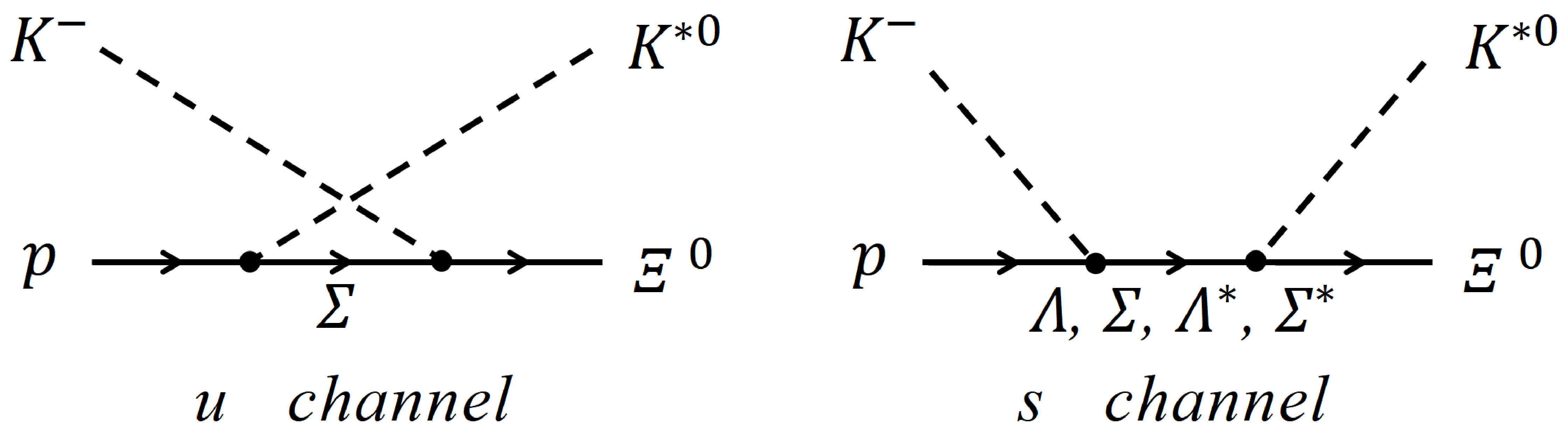}
	\caption{
		Feynman diagrams for the
		$K^-p\to K^{*0}\Xi^0$ reaction.
	}
	\label{fig2}
\end{figure}

\begin{figure}[t]
	\centering
	\includegraphics[width=0.98\columnwidth]{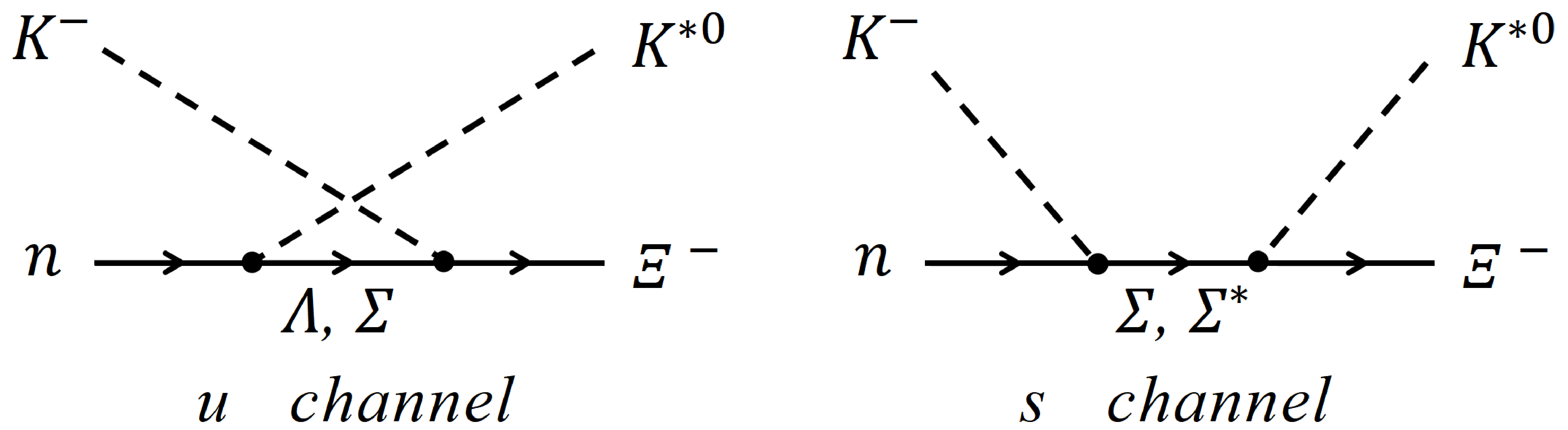}
	\caption{
		Feynman diagrams for the
		$K^-n\to K^{*0}\Xi^-$ reaction.
	}
	\label{fig3}
\end{figure}

In addition to the $K^-p\to K^{*+}\Xi^-$ reaction, we also consider the other two related charge channels
$K^-p\to K^{*0}\Xi^0$ and $K^-n\to K^{*0}\Xi^-$ within the same framework. Their tree-level Feynman diagrams are shown in \cref{fig2,fig3}, respectively. The corresponding amplitudes are obtained from
Eqs.~\eqref{eq:amplitude} to \eqref{19 amp:sSigma2230}, with the allowed intermediate hyperons determined by charge conservation, and can be written as


\begin{align}
\mathcal A^\mu_{K^-p\to K^{*0}\Xi^0}
={}&
2\mathcal A^\mu_{u,\Sigma}
+\mathcal A^\mu_{s,\Lambda}
-\mathcal A^\mu_{s,\Sigma}
\nonumber\\
&+
e^{i\phi_{\Lambda^*}}
\mathcal A^\mu_{s,\Lambda(2325)}
-
e^{i\phi_{\Sigma^*}}
\mathcal A^\mu_{s,\Sigma(2230)},
\label{eq:KpKstar0Xi0}
\end{align}
and
\begin{align}
\mathcal A^\mu_{K^-n\to K^{*0}\Xi^-}
={}&
\mathcal A^\mu_{u,\Lambda}
-\mathcal A^\mu_{u,\Sigma}
+2\mathcal A^\mu_{s,\Sigma}
\nonumber\\
&+
2e^{i\phi_{\Sigma^*}}
\mathcal A^\mu_{s,\Sigma(2230)}.
\label{eq:KnKstar0Xim}
\end{align}
Here, the relative signs and numerical coefficients arise from the corresponding isospin factors.
In particular, the initial $K^-n$ state has
total isospin $I=1$, so only $\Sigma$-type
hyperons contribute to the $s$-channel amplitude.

With the reaction amplitudes specified above, we define the
cross section and polarization observables used in the analysis.
The cross section describes the reaction strength and its kinematic
dependence, whereas the $K^*$ spin-density matrix elements and the
$\Xi$ recoil polarization characterize the spin structure of the
final state.
In the center-of-mass (c.m.) frame, the differential cross section
is given by
\begin{equation}
\frac{d\sigma}{d\Omega}
=
\frac{1}{64\pi^2s}
\frac{
\left|\bm{p}_3^{\,\mathrm{c.m.}}\right|
}{
\left|\bm{p}_1^{\,\mathrm{c.m.}}\right|
}
\left[
\frac{1}{2}
\sum_{\lambda,s_\Xi,s_N}
\left|\mathcal M_{\lambda s_\Xi s_N}\right|^2
\right].
\label{differential-cross section}
\end{equation}
The momenta $\bm{p}_1^{\,\mathrm{c.m.}}$ and
$\bm{p}_3^{\,\mathrm{c.m.}}$ refer to the incident $K^-$
and outgoing $K^*$, respectively.
The scattering angle $\theta$ is measured between these two momenta,
and the factor $1/2$ accounts for averaging over the initial nucleon spin.
The spin-density matrix of the final $K^*$ is defined as  \cite{Schilling:1970,Chen:2020Kstar,Wang:2026KstarSigmaSDME}
\begin{equation}
\rho_{\lambda\lambda'}
=
\frac{
\displaystyle
\sum_{s_N,s_\Xi}
\mathcal M_{\lambda s_\Xi s_N}
\mathcal M^*_{\lambda' s_\Xi s_N}
}{
\displaystyle
\sum_{\lambda'',s_N,s_\Xi}
\left|\mathcal M_{\lambda'' s_\Xi s_N}\right|^2
},
\quad
\mathrm{Tr}\rho=1.
\label{Kstar-SDME}
\end{equation}
The indices $\lambda,\lambda'=+1,0,-1$ label the
$K^*$ helicity states.
We consider $\rho_{00}$, $\operatorname{Re}\rho_{10}$,
and $\rho_{1-1}\equiv\operatorname{Re}\rho_{+1,-1}$.
The diagonal element $\rho_{00}$ gives the helicity-zero
population, while the off-diagonal elements describe
coherence between different helicity states.
For the recoil polarization of the final $\Xi$, we choose
the incident $K^-$ along the $+z$ axis, with the outgoing
$K^*$ lying in the $xz$ plane and having a positive
$x$ component.
The normal to the production plane is defined by
\begin{equation}
\hat{\bm n}_\Xi
=
\frac{
\bm p_K\times\bm p_\Xi
}{
\left|\bm p_K\times\bm p_\Xi\right|
},
\label{Xi-normal-direction}
\end{equation}
where $\bm p_K$ and $\bm p_\Xi$ are the c.m. three-momenta
of the incident $K^-$ and final $\Xi$, respectively.
This convention gives $\hat{\bm n}_\Xi=-\hat{\bm y}$.
The recoil polarization along this direction is \cite{Jackson:2014KXi,Luo:2026KXi}
\begin{equation}
P_\Xi
=
\frac{
\displaystyle
2\sum_{\lambda,s_N}
\operatorname{Im}
\left[
\mathcal M_{\lambda\uparrow s_N}
\mathcal M^*_{\lambda\downarrow s_N}
\right]
}{
\displaystyle
\sum_{\lambda,s_\Xi,s_N}
\left|\mathcal M_{\lambda s_\Xi s_N}\right|^2
}.
\label{Xi-recoil-polarization}
\end{equation}
The arrows $\uparrow$ and $\downarrow$ denote the spin
projections of the recoil $\Xi$ along the fixed $z$ axis.

	\section{NUMERICAL RESULTS AND DISCUSSION}
\label{zhangjie3}
	\begin{figure*}[t]
	\centering
    \hspace*{-0.3cm}
	\includegraphics[width=0.89\textwidth]{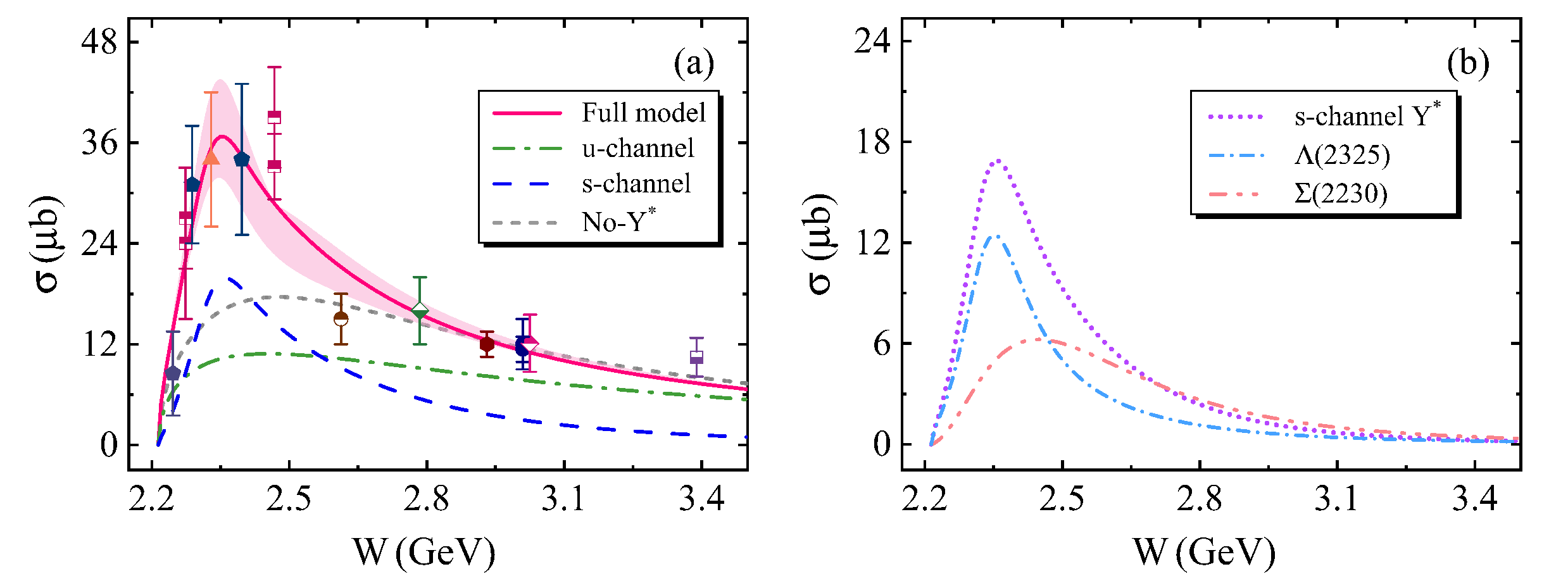}
	\caption{
    Total cross section for the
	$K^-p\rightarrow K^{*+}\Xi^-$ reaction.  (a) The full-model result and its $u$- and $s$-channel contributions, together with the No-$Y^*$ fit.
    (b) The coherent $s$-channel $Y^*$ contribution and the individual contributions from $\Lambda(2325)$ and $\Sigma(2230)$.
	The experimental data in (a) are taken from Refs.~\cite{Badier:1966KstarXi,Haque:1966KstarXi,Abrams:1968KstarXi,Goldwasser:1970Xi,Badier:1972Xi,Mazzucato:1981Xi,Bellefon:1975Xi,Dauber:1969Xi,London:1966KstarXi}, with different symbols representing different measurements.
	The shaded band in (a) represents the uncertainty propagated from the seven fitted parameters listed in \cref{tab2}.
	}
	\label{fig4}
\end{figure*}
Using the formalism developed above, we investigate the
$K^-N\to K^*\Xi$ reactions numerically.
The available cross section and polarization data for
$K^-p\to K^{*+}\Xi^-$ are first used to determine
the model parameters.
The resulting parameter set is then applied to the related
$K^-p\to K^{*0}\Xi^0$ and $K^-n\to K^{*0}\Xi^-$ reactions
without introducing additional free parameters,
and the predicted cross sections are compared with
the available experimental data.
\subsection{Fit to the $K^-p\to K^{*+}\Xi^-$ reaction}
The available experimental data on total and differential cross sections,
together with polarization measurements, are used in a combined fit
for the $K^-p\to K^{*+}\Xi^-$ reaction
\cite{Badier:1966KstarXi,Haque:1966KstarXi,
Abrams:1968KstarXi,Goldwasser:1970Xi,Badier:1972Xi,
Mazzucato:1981Xi,Bellefon:1975Xi,Dauber:1969Xi,
London:1966KstarXi}.
The free parameters are determined by minimizing $\chi^2$ using
Minuit2, and the fitted parameter values and fit quality are
summarized in \cref{tab2}.
To assess the contributions of the excited hyperons, we also
remove the $\Lambda(2325)$ and $\Sigma(2230)$ amplitudes and
refit the remaining parameters.
This alternative is referred to as the No-$Y^*$ fit.
The full model yields $\chi^2/\mathrm{d.o.f.}=2.32$,
whereas the corresponding value increases to $3.98$
for the No-$Y^*$ fit, with the deterioration in fit quality
arising mainly from the cross section data.
\begin{table}[h]
    \renewcommand{\arraystretch}{1.5}
    \caption{Fitted parameters and fit quality for the
    $K^-p\to K^{*+}\Xi^-$ reaction using the available experimental
		data in Refs.~\cite{Badier:1966KstarXi,Haque:1966KstarXi,
		Abrams:1968KstarXi,Goldwasser:1970Xi,Badier:1972Xi,
		Mazzucato:1981Xi,Bellefon:1975Xi,Dauber:1969Xi,London:1966KstarXi}.}
    \label{tab2}
    \centering
    {\footnotesize
    \setlength{\tabcolsep}{4mm}
    \begin{tabular}{ccc}
        \hline\hline
        Parameter & Full model fit & No-$Y^*$ fit \\
        \hline
        $\Lambda_u$ (GeV)
            & $0.475\pm0.005$ & $0.483\pm0.003$ \\
        $\Lambda_{s,Y}$ (GeV)
            & $0.784\pm0.009$ & $0.802\pm0.009$ \\
        $\Lambda_{s,Y^*}$ (GeV)
            & $1.670\pm0.314$ & --\\
        $g^1_{K^*\Xi\Lambda(2325)}$
            & $0.457\pm0.116$ & -- \\
        $g^1_{K^*\Xi\Sigma(2230)}$
            & $1.845\pm0.945$ & -- \\
        $\phi_{\Lambda^*}$ ($^\circ$)
            & $100.4\pm31.0$ & -- \\
     
        $\phi_{\Sigma^*}$ ($^\circ$)
            & $-81.1\pm17.3$ & -- \\
        \hline
        $\chi^2/\mathrm{d.o.f.}$
            & $2.32$ & $3.98$ \\
        \hline\hline
    \end{tabular}
    }
\end{table}

\cref{fig4}(a) shows the total cross section for the $K^-p\to K^{*+}\Xi^-$ reaction.
The full model provides a good description of the overall energy dependence of the experimental data.
The cross section rises rapidly above the $K^*\Xi$ threshold, reaches a peak around
$W\simeq2.35$--$2.45$~GeV, and then gradually
decreases at higher energies.
This peak is mainly associated with the $s$-channel contribution, whereas the $u$-channel contribution exhibits a smoother energy dependence and becomes relatively more important at higher energies.
By contrast, the No-$Y^*$ fit yields a substantially smaller cross section near threshold and fails to reproduce the pronounced peak observed in the experimental data.
\cref{fig4}(b) further compares the combined coherent contribution of the $s$-channel excited hyperons with their individual contributions.
The results show that both $\Lambda(2325)$ and $\Sigma(2230)$ contribute significantly to the near-threshold peak, highlighting the sensitivity of the $K^-p\to K^{*+}\Xi^-$ reaction to high-mass hyperon excitations.

\begin{figure}[t]
    \centering
    \includegraphics[width=0.92\columnwidth]{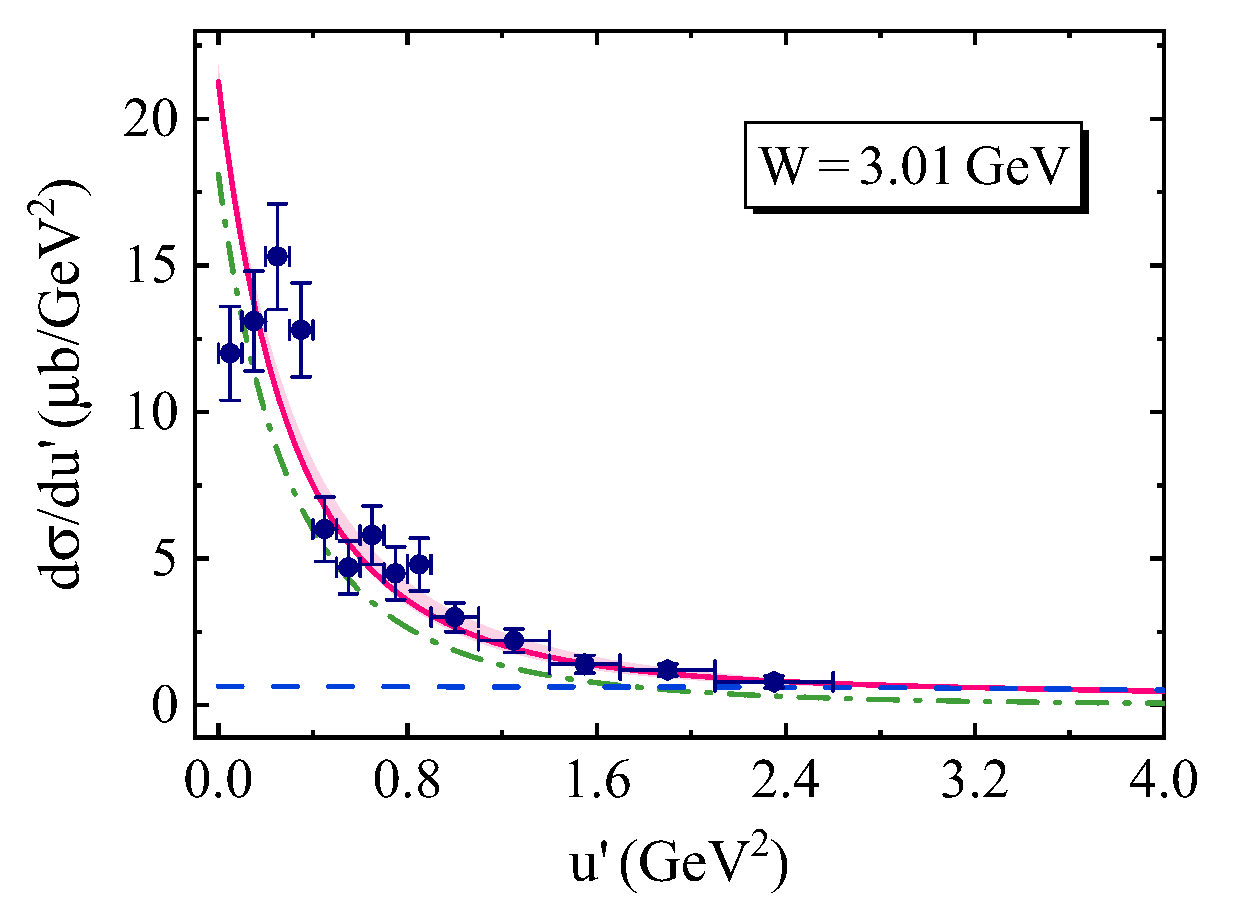}
    \caption{Differential cross section as a function of the momentum-transfer variable $u'$ for the $K^-p\to K^{*+}\Xi^-$ reaction at $W=3.01$ GeV.
    The experimental data are taken from
    Ref.~\cite{Mazzucato:1981Xi}.
    The line styles are the same as in \cref{fig4}(a).}
    \label{fig5}
\end{figure}

\begin{figure}[t]
    \centering
    \includegraphics[width=0.92\columnwidth]{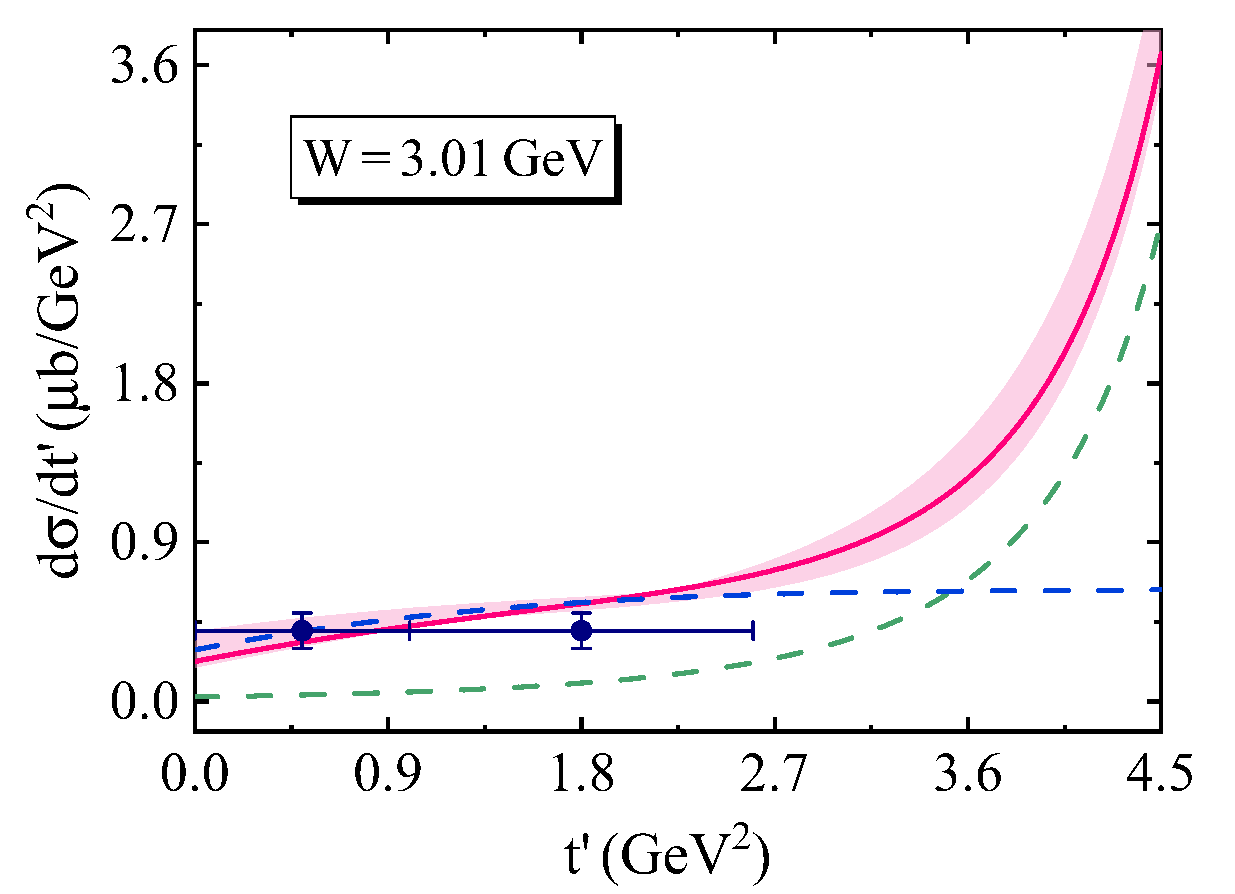}
    \caption{Differential cross section as a function of the momentum-transfer variable $t'$ for the $K^-p\to K^{*+}\Xi^-$ reaction at $W=3.01$ GeV.
    The experimental data are taken from
    Ref.~\cite{Mazzucato:1981Xi}.
    The line styles are the same as in \cref{fig4}(a).}
    \label{fig6}
\end{figure}

\begin{figure}[t]
    \centering
    \includegraphics[width=0.92\columnwidth]{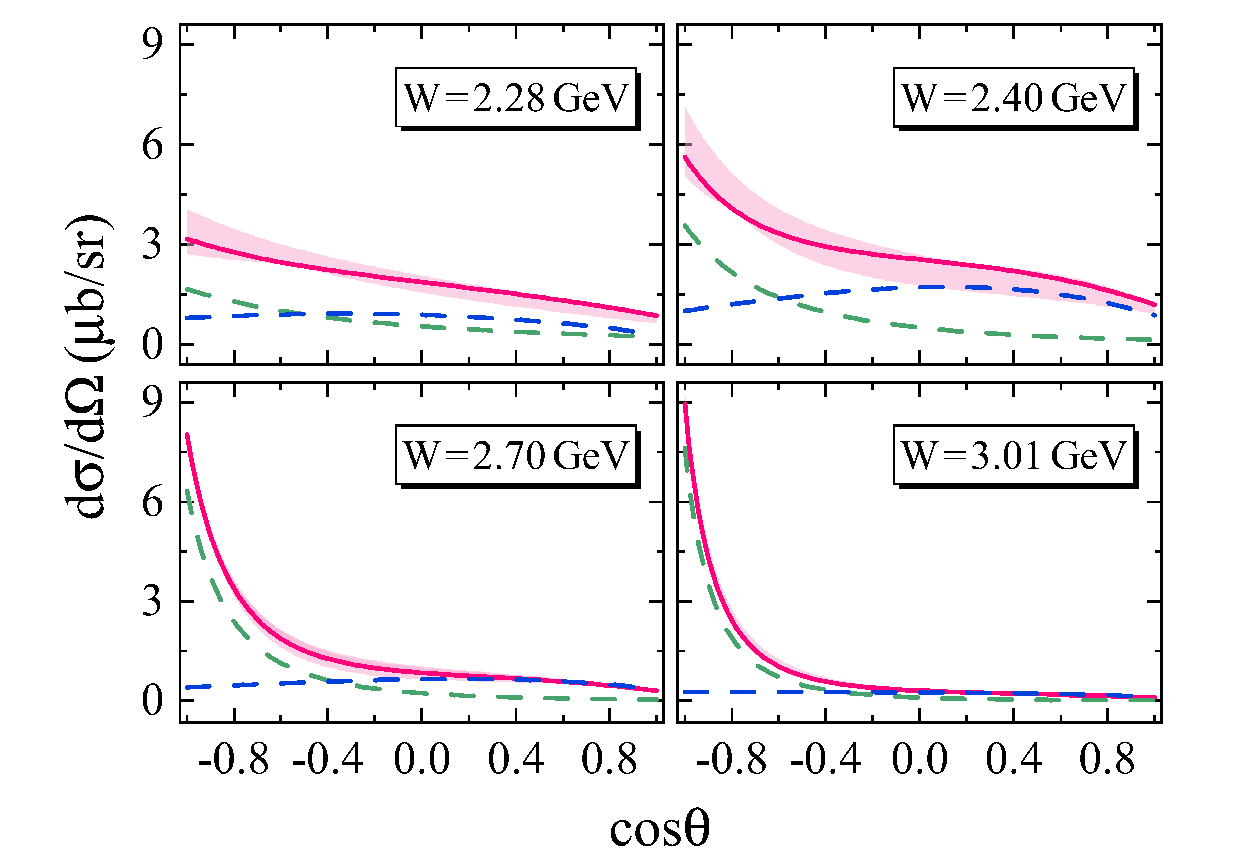}
    \caption{Predicted angular distributions of the
    differential cross section for the
    $K^-p\to K^{*+}\Xi^-$ reaction at
    $W=2.28$, $2.40$, $2.70$, and $3.01$~GeV.
    The line styles are the same as in \cref{fig4}(a).}
    \label{fig7}
\end{figure}

\Cref{fig5,fig6} show the momentum-transfer
distributions in $u'$ and $t'$ for the $K^-p\to K^{*+}\Xi^-$ reaction
at $W=3.01$ GeV, respectively, where $u'=|u-u_{\max}|$
and $t'=|t-t_{\max}|$.
The $u'$ distribution in \cref{fig5} falls rapidly
with increasing $u'$, indicating that the cross
section is concentrated in the small-$u'$ region.
In contrast, the $t'$ distribution in \cref{fig6}
exhibits a relatively weak variation over the
measured range.
At larger $t'$, the calculated $u$-channel
contribution becomes increasingly important
and eventually dominates. To further investigate the angular dependence
of the reaction, \cref{fig7} presents the
differential cross sections at several c.m. energies.
The angular distributions exhibit a clear evolution
with increasing energy.
At $W=2.28$ and $2.40$~GeV, the distributions
are relatively broad, with appreciable
contributions from the $s$-channel.
As the energy increases, the backward-angle
enhancement becomes more pronounced.
At $W=2.70$ and $3.01$~GeV, the distributions
become strongly peaked at backward angles,
primarily due to $u$-channel exchange.
Overall, the near-threshold peak in the total
cross section is mainly associated with the
$s$-channel excited-hyperon contributions,
whereas the differential cross sections at
higher energies increasingly exhibit the
characteristic behavior of $u$-channel exchange.
The total and differential cross sections thus
provide complementary information on the
underlying reaction mechanisms, and measurements
of angular distributions at different energies
would help further constrain the relative
contributions of the $s$- and $u$-channel exchanges.

\begin{figure}[t]
    \centering
    \hspace*{-0.15cm}
   \includegraphics[width=0.496\textwidth]{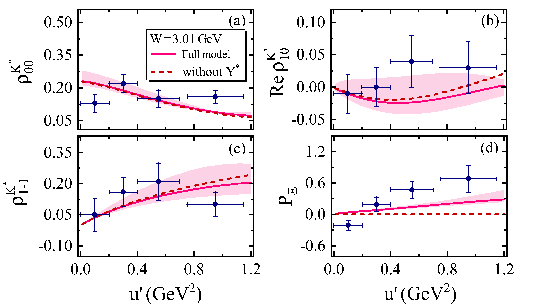}
    \caption{$K^*$ spin-density matrix elements $\rho_{00}$, $\operatorname{Re}\rho_{10}$, and $\rho_{1-1}$, together with the $\Xi$ recoil polarization $P_\Xi$, as functions of the momentum-transfer variable $u'$ for the $K^-p\to K^{*+}\Xi^-$ reaction at $W=3.01$ GeV. The experimental data are taken from Ref.~\cite{Mazzucato:1981Xi}.
    The two theoretical curves correspond to the full model and the calculation with the $Y^*$ amplitudes switched off.
    In the latter case, the remaining parameters are kept at their full-model values without refitting.}
    \label{fig8}
\end{figure}

\cref{fig8} shows the $u'$ dependence of the $K^*$ spin-density matrix elements $\rho_{00}$,
$\mathrm{Re}\rho_{10}$, and $\rho_{1-1}$, as well as the $\Xi$ recoil polarization $P_\Xi$,
for the $K^-p\to K^{*+}\Xi^-$ reaction at $W=3.01$~GeV.
As seen from the cross section results in \cref{fig4}(a),
the $s$-channel contribution to the cross section is
relatively small at this energy.
Correspondingly, with all other parameters kept fixed, the differences between the full-model results and those obtained by switching off the $Y^*$ amplitudes are generally
small for the three spin-density matrix elements.
In contrast, $P_\Xi$ still exhibits a pronounced difference: 
the full model yields a nonzero recoil polarization that increases with $u'$, whereas $P_\Xi$ approaches zero when the $Y^*$ amplitudes are switched off.
This indicates that the recoil polarization remains sensitive to interference effects involving the excited-hyperon amplitudes.
Nevertheless, at larger $u'$, the full-model predictions remain below the central values of the experimental data.

\begin{figure}[t]
    \centering
    \hspace*{-0.15cm}
   \includegraphics[width=0.496\textwidth]{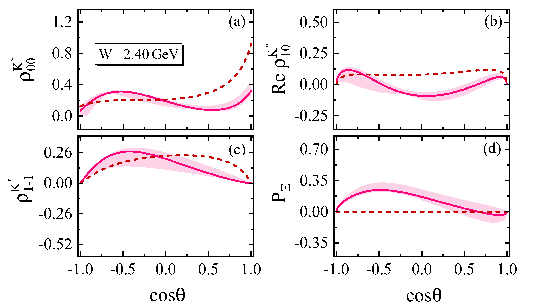}
   \caption{Predicted angular distributions of
   $\rho_{00}$, $\operatorname{Re}\rho_{10}$,
   $\rho_{1-1}$, and $P_\Xi$ as functions of
   $\cos\theta$ for the $K^-p\to K^{*+}\Xi^-$
   reaction at $W=2.40$~GeV.
   The line styles are the same as in \cref{fig8}.}
    \label{fig9}
\end{figure}

\begin{figure*}[t]
    \centering
    \hspace*{-0.3cm}
    \includegraphics[width=0.89\textwidth]{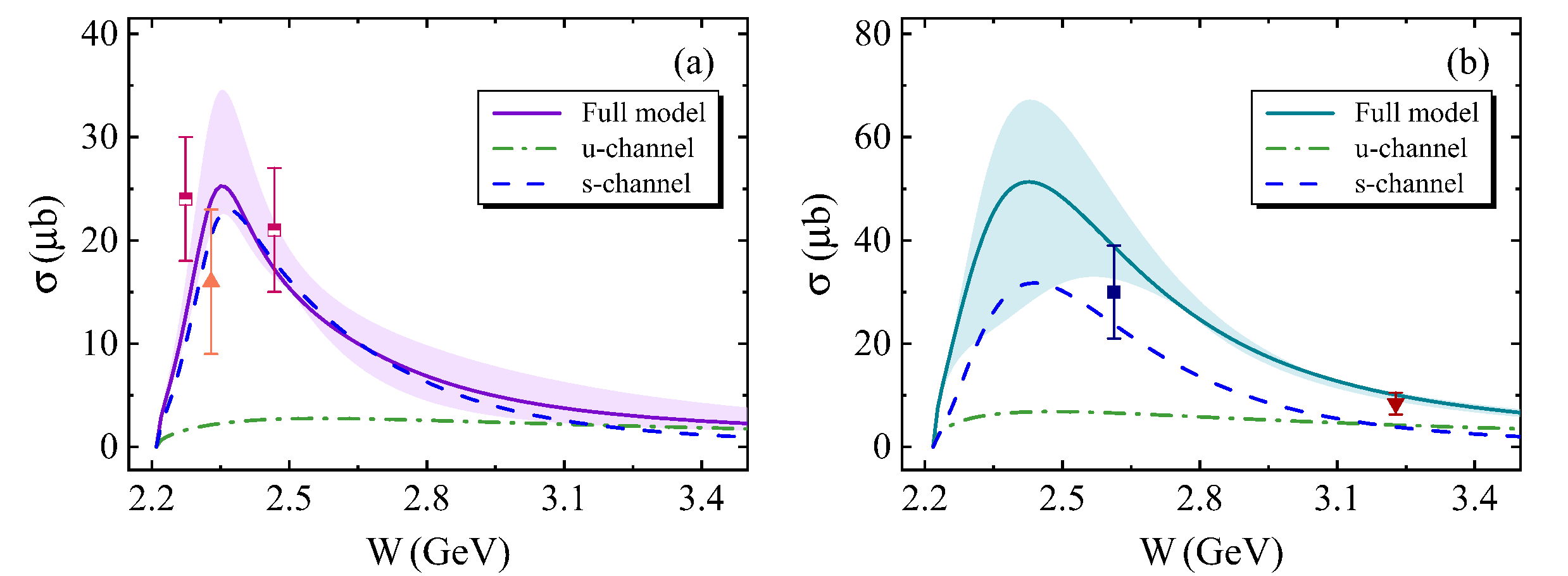}
   \caption{Total cross sections for the
   $K^-p\to K^{*0}\Xi^0$ (a) and
   $K^-n\to K^{*0}\Xi^-$ (b) reactions.
   The experimental data are taken from
   Refs.~\cite{Dauber:1969Xi,London:1966KstarXi,
   Scheuer:1971Kn,DiBianca:1975XiOmega},
   with different symbols representing different
   measurements.}
    \label{fig10}
\end{figure*}

To further investigate the polarization observables near threshold, \cref{fig9} presents the predicted angular distributions of $\rho_{00}$, $\operatorname{Re}\rho_{10}$, $\rho_{1-1}$, and $P_\Xi$ for the $K^-p\to K^{*+}\Xi^-$ reaction at $W=2.40$~GeV.
All four observables exhibit pronounced angular dependence in the full model.
Compared with the results obtained by switching off the $Y^*$ amplitudes, $\rho_{00}$ shows significant differences at forward angles, while $\operatorname{Re}\rho_{10}$ exhibits opposite signs in certain angular regions.
The peak position of $\rho_{1-1}$ also shifts noticeably, and the full model predicts a pronounced nonzero peak in $P_\Xi$ at backward
angles.
Compared with the results at $W=3.01$~GeV in \cref{fig8}, the three spin-density matrix elements exhibit more pronounced differences between the two calculations in certain angular regions near threshold, while $P_\Xi$ remains strongly sensitive to the excited-hyperon contributions.
Therefore, measurements of polarization angular distributions near threshold could provide valuable complementary information for constraining the contributions of excited hyperons and understanding the underlying reaction mechanisms.

\subsection{Predictions for the $K^-p\to K^{*0}\Xi^0$ and $K^-n\to K^{*0}\Xi^-$ reactions}

Based on the isospin relations given above, we further apply
the theoretical framework to the two related reaction channels
$K^-p\to K^{*0}\Xi^0$ and $K^-n\to K^{*0}\Xi^-$.
Using the best-fit parameters of the full model listed in
Table~II, we predict their cross sections without introducing
additional free parameters and compare the results with the
available experimental data to assess the model's predictive
capability across different reaction channels.

\begin{figure}[t]
    \centering
      \hspace*{-0.15cm}
    \includegraphics[width=0.92\columnwidth]{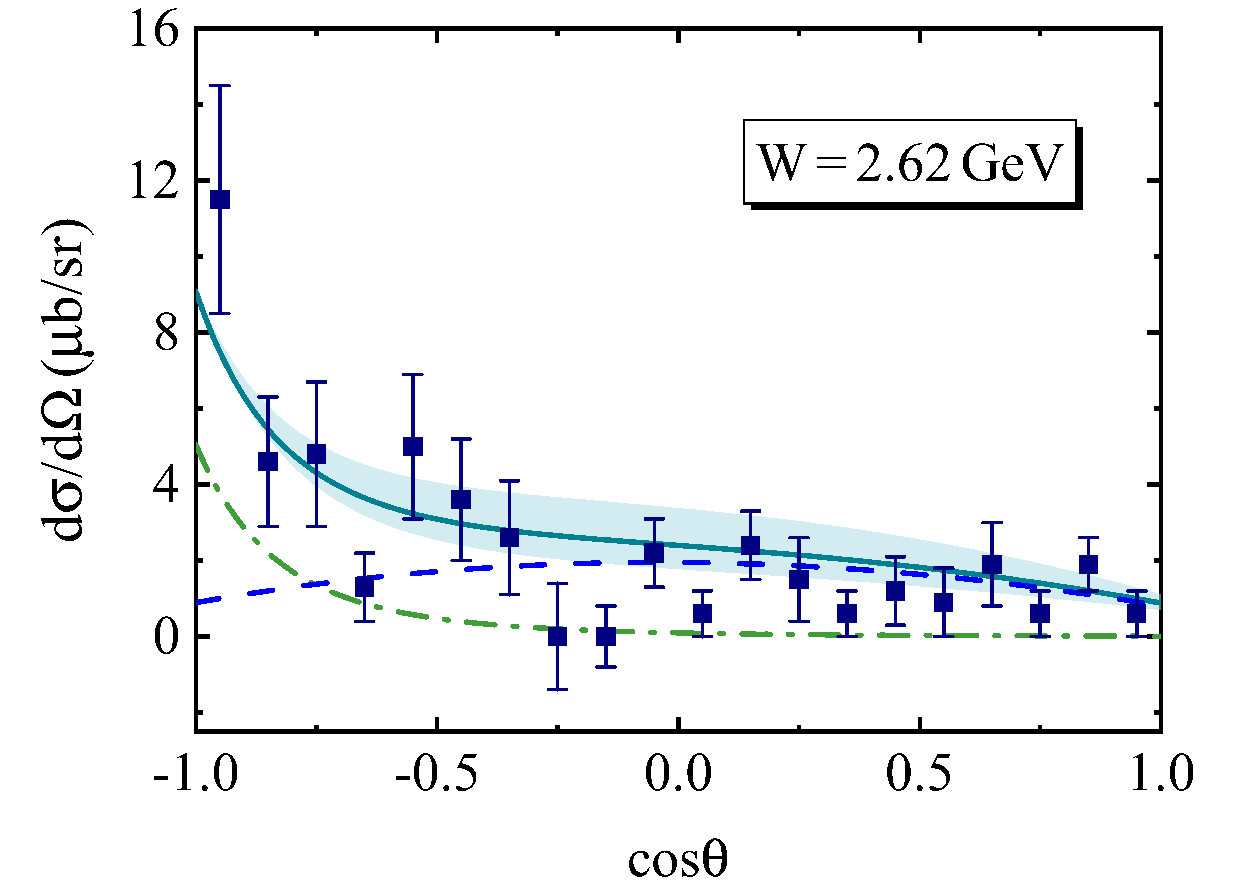}
    \caption{Differential cross section as a function of $\cos\theta$ for the
    $K^-n\to K^{*0}\Xi^-$ reaction at
    $W=2.62$~GeV.
    The experimental data are taken from
    Ref.~\cite{Scheuer:1971Kn}.
    The curve notation is the same as in \cref{fig10}(b).}
    \label{fig11}
\end{figure}

\cref{fig10}(a) and (b) show the predicted total cross sections for the $K^-p\to K^{*0}\Xi^0$ and $K^-n\to K^{*0}\Xi^-$ reactions, respectively.
Both reactions exhibit pronounced peaks near threshold, where the $s$-channel contributions dominate.
As the energy increases, the relative importance of the $u$-channel contributions gradually grows.
Although the overall energy dependences are similar, the two channels differ noticeably in their cross section magnitudes and the relative contributions of the exchange mechanisms.
These differences reflect the allowed intermediate-hyperon exchanges and the corresponding isospin factors in each channel.
\cref{fig11} presents the angular distribution for the $K^-n\to K^{*0}\Xi^-$ reaction at $W=2.62$~GeV, together with the available experimental data.
The theoretical result provides a good description of the backward-angle enhancement observed in the data, which is driven mainly by $u$-channel exchange.
The comparisons with the experimental data in \cref{fig10,fig11} indicate that the isospin relations and the common set of model parameters used here yield predictions that are generally consistent with the measurements across different reaction channels.

\begin{figure}[h]
    \centering
     \hspace*{-0.15cm}
    \includegraphics[width=0.92\columnwidth]{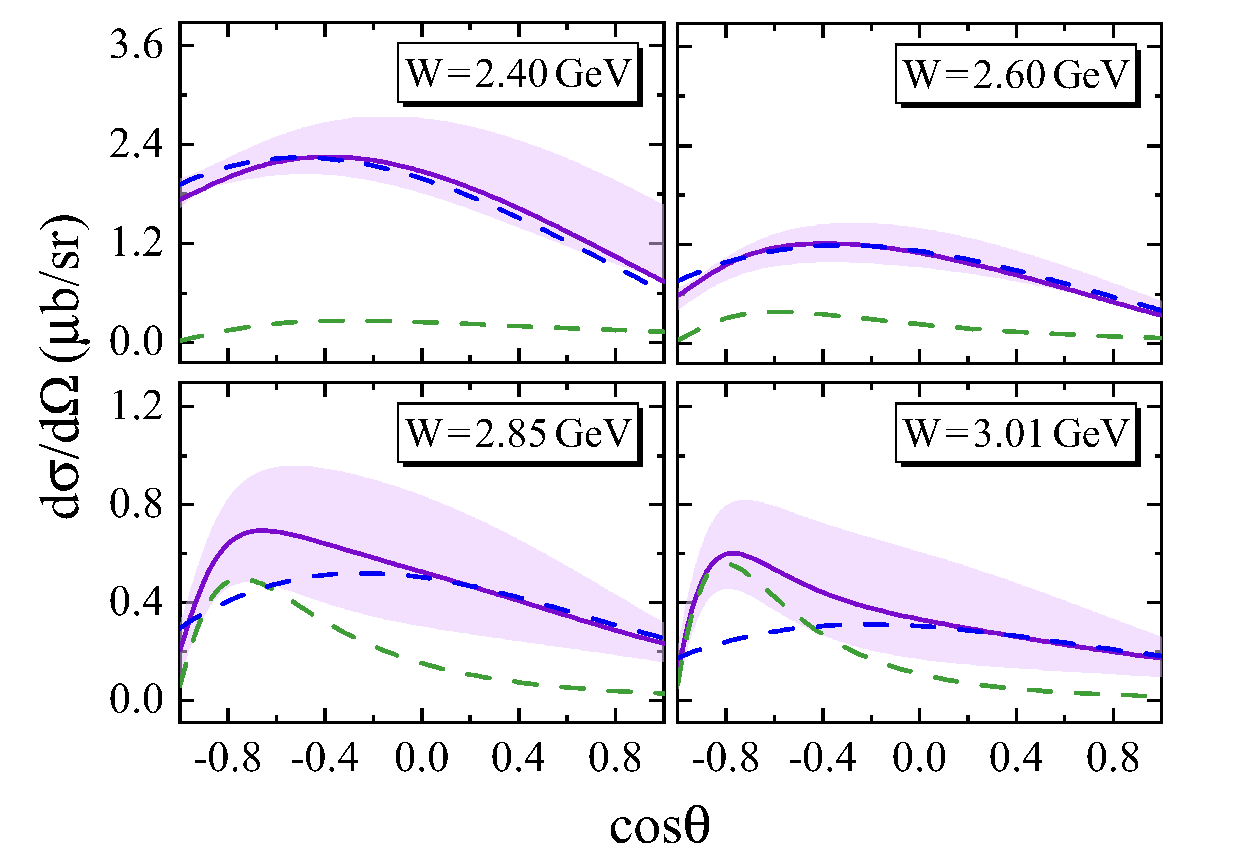}
    \caption{Predicted angular distributions of the differential cross section for the $K^-p\to K^{*0}\Xi^0$ reaction at $W=2.40$, $2.60$, $2.85$, and $3.01$~GeV. The curve notation is the same as in \cref{fig10}(a).}
    \label{fig12}
\end{figure}
\begin{figure}[!t]
    \centering
    \hspace*{-0.15cm}
    \includegraphics[width=0.92\columnwidth]{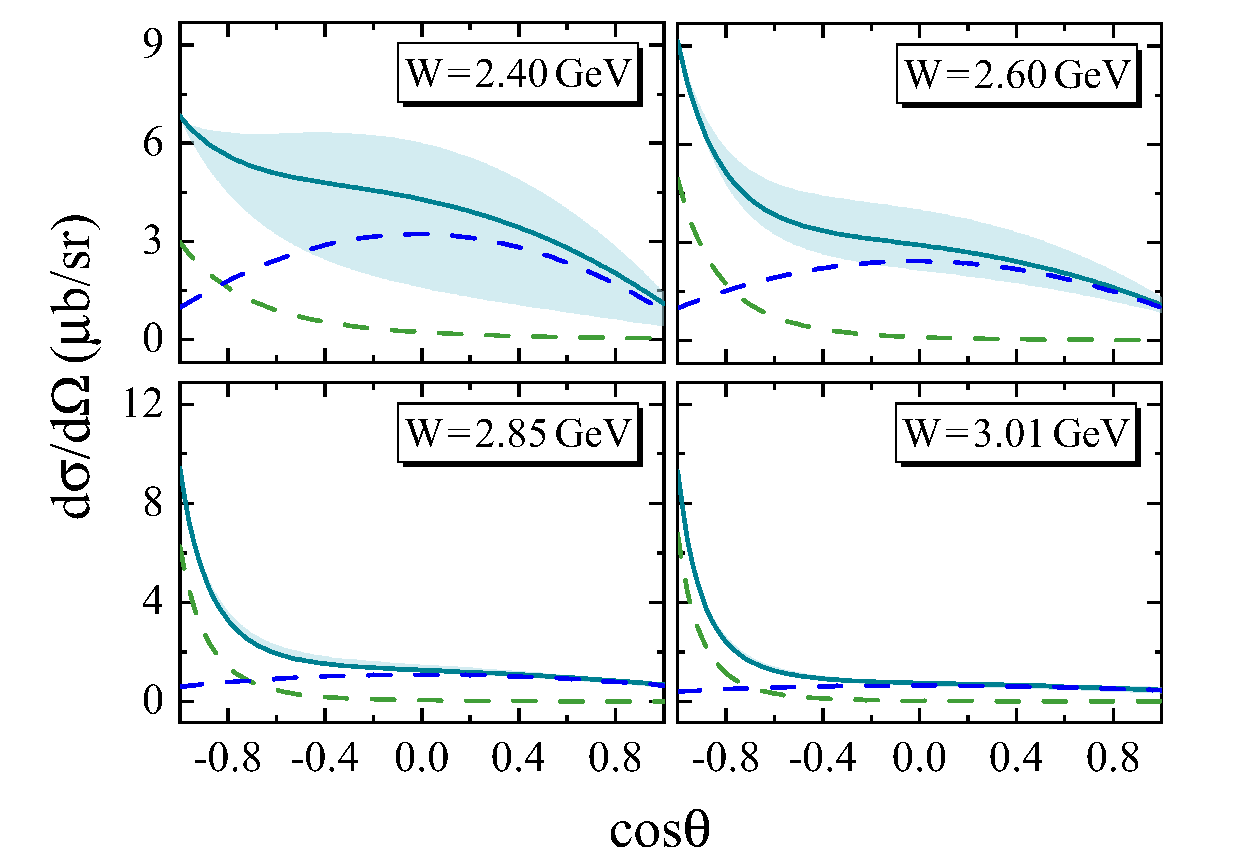}
    \caption{Predicted angular distributions of the differential cross section for the $K^-n\to K^{*0}\Xi^-$ reaction at $W=2.40$, $2.60$, $2.85$, and $3.01$~GeV. The curve notation is the same as in \cref{fig10}(b).}
    \label{fig13}
\end{figure}

\cref{fig12,fig13} present the predicted angular
distributions for the $K^-p\to K^{*0}\Xi^0$
and $K^-n\to K^{*0}\Xi^-$ reactions,
respectively, at several c.m. energies.
Both channels exhibit an energy evolution similar to that of the $K^-p\to K^{*+}\Xi^-$ reaction shown in \cref{fig7}, with the backward-angle enhancement becoming more pronounced as the energy increases.
This enhancement is more prominent in the
$K^-n\to K^{*0}\Xi^-$ channel.

\begin{figure}[t]
    \centering
    \hspace*{-0.15cm}
    \includegraphics[width=0.496\textwidth]{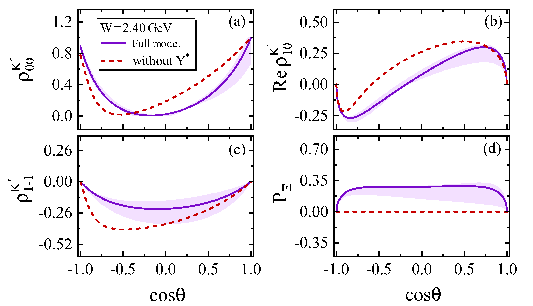}
   \caption{Predicted angular distributions of $\rho_{00}$, $\operatorname{Re}\rho_{10}$, $\rho_{1-1}$, and $P_\Xi$ as functions of $\cos\theta$ for the $K^-p\to K^{*0}\Xi^0$ reaction at $W=2.40$~GeV. The solid and dashed curves represent the full model and the calculation without $Y^*$, respectively.}
    \label{fig14}
\end{figure}
\begin{figure}[t]
    \centering
   \hspace*{-0.15cm}
   \includegraphics[width=0.496\textwidth]{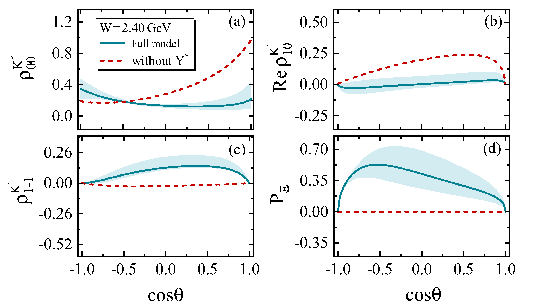}
    \caption{Predicted angular distributions of $\rho_{00}$, $\operatorname{Re}\rho_{10}$, $\rho_{1-1}$, and $P_\Xi$ as functions of $\cos\theta$ for the $K^-n\to K^{*0}\Xi^-$ reaction at $W=2.40$~GeV. The solid and dashed curves represent the full model and the calculation without $Y^*$, respectively.}
    \label{fig15}
\end{figure}

To further investigate the spin structure of different reaction
channels, \cref{fig14,fig15} present predictions for the angular
distributions of $\rho_{00}$, $\mathrm{Re}\rho_{10}$, $\rho_{1-1}$,
and $P_\Xi$ for the $K^-p\to K^{*0}\Xi^0$ and
$K^-n\to K^{*0}\Xi^-$ reactions, respectively, at $W=2.40$~GeV.
As in the $K^-p\to K^{*+}\Xi^-$ results at the same energy
shown in \cref{fig9}, the polarization angular distributions
in both channels change significantly when the $Y^*$
amplitudes are switched off.
Notably, for the $K^-n\to K^{*0}\Xi^-$ reaction, the
$s$-channel excited-hyperon contribution comes exclusively
from $\Sigma(2230)$.
Therefore, polarization measurements in this channel are
expected to provide additional constraints beyond cross section
data and help clarify the role of $\Sigma(2230)$ in
$K^*\Xi$ production.

At present, experimental data for these two reaction channels
remain limited.
Precise near-threshold measurements, as well as
angular-distribution and polarization data at different
energies, are particularly scarce.
More precise total cross section measurements near threshold
would help test the peak structures predicted in this work,
while differential cross section measurements over a wider
energy range, particularly at backward angles, would further
test the energy evolution of the angular distributions.
In addition, polarization measurements would provide
spin-dependent information beyond that available from
cross section data and further constrain the relative
contributions of excited hyperons and other exchange mechanisms,
thereby providing a more comprehensive test of the model's
description of the production mechanisms in different
reaction channels.

	\section{Dalitz process and experimental feasibility}\label{zhangjie4}

\begin{figure}[h]
    \centering
    \includegraphics[width=0.95\columnwidth]{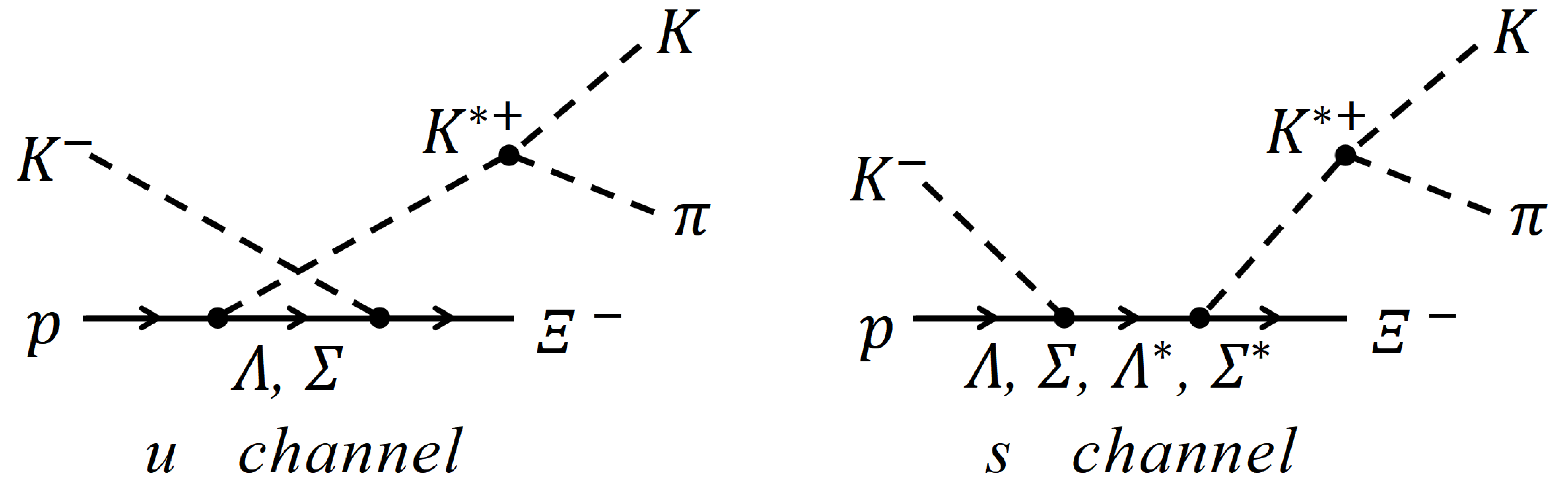}
    \caption{Feynman diagrams for
    $K^-p\to K^{*+}\Xi^-\to K\pi\Xi^-$,
    where $K\pi$ denotes the sum of the
    $K^+\pi^0$ and $K^0\pi^+$ modes.}
    \label{fig16}
\end{figure}

In experiments, the short-lived $K^{*+}$ meson produced in the $K^-p\to K^{*+}\Xi^-$ reaction must be reconstructed from its decay products. Since the branching fraction for $K^{*+}\to K\pi$ is nearly $100\%$~\cite{ParticleDataGroup:2026}, we further consider the $K^-p\to K^{*+}\Xi^-\to K\pi\Xi^-$ process to examine the possibility of reconstructing $K^{*+}\Xi^-$ production through the $K\pi\Xi^-$ final state. Here, $K\pi$ denotes the sum over the two allowed charge modes, $K^+\pi^0$ and $K^0\pi^+$. The corresponding Feynman diagrams are shown in \cref{fig16}.

The $K\pi$ invariant-mass distribution for the $K^-p\to K^{*+}\Xi^-\to K\pi\Xi^-$ process can be expressed in terms of the corresponding two-body $K^{*+}\Xi^-$ production cross section as \cite{Wang:2024KstarSigma,Gao:2026Lambda,Wang:2024qnk,Kim:2017nxg}
\begin{equation}
	\begin{aligned}[b]
		\frac{
			\mathrm{d}\sigma_{K^-p\rightarrow K^{*+}\Xi^-
			\rightarrow K\pi\Xi^-}
		}{
			\mathrm{d}M_{K\pi}
		}
		&\simeq
		\frac{2M_{K^{*+}}M_{K\pi}}{\pi}
		\\
		&\quad\times
		\frac{
			\sigma_{K^-p\rightarrow K^{*+}\Xi^-}
			\Gamma_{K^{*+}\rightarrow K\pi}
		}{
			\left(M_{K\pi}^{2}-M_{K^{*+}}^{2}\right)^{2}
			+M_{K^{*+}}^{2}\Gamma_{K^{*+}}^{2}
		}.
	\end{aligned}
	\label{eq:dalitz}
\end{equation}
Here, $M_{K\pi}$ is the invariant mass of the $K\pi$ system,
while $M_{K^{*+}}$ and $\Gamma_{K^{*+}}$ denote the mass and total width
of the $K^{*+}$, respectively.
In the calculation, we take
$\Gamma_{K^{*+}\to K\pi}\simeq\Gamma_{K^{*+}}=48.5$~MeV
\cite{ParticleDataGroup:2026}.

\begin{figure}[t]
    \centering
    \includegraphics[width=0.92\columnwidth]{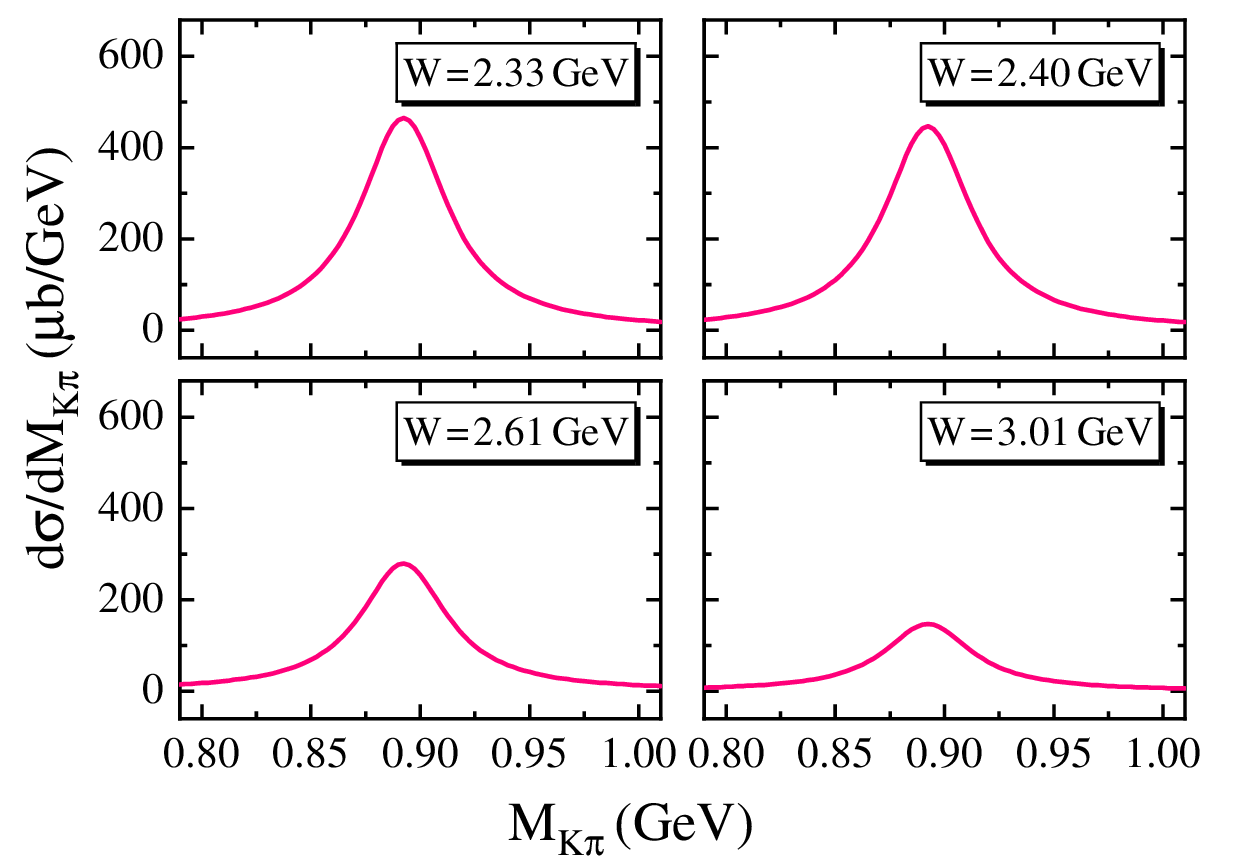}
    \caption{The $K\pi$ invariant-mass distribution
    $\mathrm{d}\sigma/\mathrm{d}M_{K\pi}$ for
    $K^-p\to K^{*+}\Xi^-\to K\pi\Xi^-$ at
    $W=2.33$, $2.40$, $2.61$, and $3.01$~GeV.}
    \label{fig17}
\end{figure}

\cref{fig17} shows the calculated $K\pi$ invariant-mass distributions at different c.m. energies. At all four energies, a clear peak appears near $M_{K\pi}\simeq 0.892$~GeV, corresponding to the $K^{*+}(892)$ resonance. The peak height varies with the c.m. energy, reflecting the energy dependence of the $K^-p\to K^{*+}\Xi^-$ production cross section. This resonance structure indicates that the $K\pi$ invariant-mass spectrum can provide a distinct signature for the experimental reconstruction of $K^{*+}$.

To further assess the experimental prospects for reconstructing the $K^{*+}(892)$ resonance through the $K\pi\Xi^-$ final state, we compare the theoretical total cross section $\sigma_2$ for $K^-p\to K^{*+}\Xi^-$ with the sum of the measured total cross sections $\sigma_1$ for the two $K\pi\Xi^-$ charge channels, as listed in \cref{tab5}. 
At the energies considered, $\sigma_2$ is appreciable relative to $\sigma_1$. In particular, the ratio $\sigma_2/\sigma_1$ is approximately $45\%$ at $W=2.397$ and $2.460$~GeV.
Together with the clear $K^{*+}(892)$ resonance peak shown in \cref{fig17}, these results suggest promising prospects for reconstructing and studying the $K^-p\to K^{*+}\Xi^-$ reaction through the $K\pi\Xi^-$ final state.

\begin{table}[t]
    \renewcommand{\arraystretch}{1.5}
    \caption{Experimental and calculated total cross sections
    at selected c.m. energies.
    The quantity $\sigma_1$ is the sum of the experimental cross sections
    for $K^-p\to K^+\pi^0\Xi^-$ and $K^-p\to K^0\pi^+\Xi^-$
    \cite{Bellefon:1977Channels,Smith:1965Xi,Merrill:1970Xi,Mazzucato:1981Xi},
    while $\sigma_2$ is the calculated total cross section
    for $K^-p\to K^{*+}\Xi^-$.}
    \label{tab5}
    \centering
    {\footnotesize
    \setlength{\tabcolsep}{10pt}
    \begin{tabular}{c|c|c|c}
        \hline\hline
        $W$ (GeV) & $\sigma_1$ ($\mu\mathrm{b}$)
        & $\sigma_2$ ($\mu\mathrm{b}$) & $\sigma_2/\sigma_1$ \\
        \hline
        2.246 & $118\pm9.4$ & 13.73 & 11.64\% \\
        \hline
        2.328 & $147\pm10$ & 35.17 & 23.93\% \\
        \hline
        2.397 & $78\pm8.6$ & 34.29 & 43.96\% \\
        \hline
        2.460 & $64.2\pm2.8$ & 29.26 & 45.58\% \\
        \hline
        2.613 & $60\pm6.4$ & 21.18 & 35.30\% \\
        \hline
        3.010 & $44.5\pm5.0$ & 11.22 & 25.21\% \\
        \hline\hline
    \end{tabular}}
\end{table}

	\section{Summary} \label{zhangjie5}

In this work, we have investigated the production mechanisms of the $K^-N \to K^*\Xi$ reactions and the role of high-mass excited hyperons within an effective Lagrangian approach. The model includes ground-state $\Lambda$ and $\Sigma$ exchanges in the $s$- and $u$-channels, together with the $s$-channel contributions from $\Lambda(2325)$ and $\Sigma(2230)$. The free parameters were determined from a combined fit to the cross section and polarization data for $K^-p \to K^{*+}\Xi^-$. The fit indicates that the experimental data require the inclusion of excited-hyperon contributions. Specifically, for the $K^-p \to K^{*+}\Xi^-$ reaction, $\Lambda(2325)$ and $\Sigma(2230)$ are essential for reproducing the near-threshold peak in the total cross section. Without these states, the model fails to reproduce this peak even after refitting the remaining parameters. Furthermore, keeping the fitted parameters fixed, the model also reproduces the available cross section data for $K^-p \to K^{*0}\Xi^0$ and $K^-n \to K^{*0}\Xi^-$. This result further supports the applicability of the adopted production mechanism across different charge channels.

Further analysis shows that the three reactions exhibit similar patterns of exchange contributions across different energy regions. Near threshold, the total cross sections are dominated by $s$-channel contributions, whereas at higher energies, $u$-channel exchanges dominate the backward-angle differential cross sections. These different energy and angular dependences provide an important basis for distinguishing the contributions from excited and ground-state hyperon exchanges. Moreover, the $K^-n \to K^{*0}\Xi^-$ reaction exhibits a particularly interesting feature: within the present model, its $s$-channel excited-hyperon contribution arises exclusively from $\Sigma(2230)$. This makes the reaction an ideal channel for investigating the role of $\Sigma(2230)$, and future experimental measurements could provide more targeted constraints on its contribution to $K^*\Xi$ production.

Cross sections alone, however, do not fully test this mechanism.
Polarization observables provide an independent and in some regions more sensitive probe of the spin structure.
For all three reactions, the predicted $K^*$ spin-density matrix elements and $\Xi$ recoil polarization are sensitive to excited-hyperon contributions near threshold.
For $K^-p\to K^{*+}\Xi^-$, the $\Xi$ recoil polarization remains sensitive to interference involving excited-hyperon amplitudes even at higher energies, where their contributions to the cross section are relatively small.
This persistence means that polarization can constrain the excited-hyperon amplitudes in a region where cross section data alone would lose sensitivity.

The Dalitz analysis provides a further avenue for experimental investigation. For the $K^-p \to K^{*+}\Xi^- \to K\pi\Xi^-$ process, the calculated $K\pi$ invariant-mass spectrum exhibits a clear $K^{*+}(892)$ resonance peak, while comparison with existing $K\pi\Xi^-$ cross section data supports the experimental feasibility of reconstructing $K^{*+}\Xi^-$ production. High-intensity beam facilities and hadron spectroscopy programs \cite{wl:2024ext}, including HIAF/HHaS \cite{Chen:2026HHaS}, J-PARC \cite{Aoki:2021cqa}, AMBER \cite{Adams:2018pwt}, and JLab \cite{Dobbs:2022KLF}, are expected to provide new experimental opportunities for strange-hadron production and spectroscopy. Future measurements of near-threshold total cross sections, differential cross sections over wider angular ranges at different energies, and the $K^*$ spin-density matrix elements and $\Xi$ recoil polarization would directly test the predicted near-threshold peak structures, backward-angle enhancements, and polarization features. A combined analysis of these observables across the three reactions would provide a more comprehensive test of the model and help constrain the relative contributions of different exchange mechanisms. Such studies could deepen our understanding of high-mass excited hyperons and their hadronic couplings, providing valuable information for further investigations of the excited-hyperon spectrum.

	\section*{ACKNOWLEDGMENTS}

This work is supported by the National Natural Science Foundation of China under Grants No.~12565018, No.~12065014, and No.~12247101, and by the Natural Science Foundation of Gansu Province under Grant No.~22JR5RA266. We also acknowledge the West Light Foundation of the Chinese Academy of Sciences under Grant No.~21JR7RA201 and the ``Innovation Star'' Program for Postgraduates of the Gansu Provincial Department of Education under Grant No.~2026CXZX-611.

    \section*{Data Availability}
No new experimental data were generated in this theoretical study. All numerical results can be reproduced using the model parameters and formulas provided in this article and are available from the authors upon reasonable request.


\begin{thebibliography}{99}



\bibitem{ParticleDataGroup:2026}
F.~Takahashi \textit{et al.} (Particle Data Group),
Review of Particle Physics,
Int. J. Mod. Phys. A \textbf{41}, 2630011 (2026).

\bibitem{Bulava:2024HadronSpectroscopy}
J.~Bulava,
Hadron spectroscopy with lattice QCD:
Challenges and opportunities,
EPJ Web Conf. \textbf{303}, 01003 (2024).

\bibitem{Klempt:2009pi}
E.~Klempt and J.~M.~Richard,
Baryon spectroscopy,
Rev. Mod. Phys. \textbf{82}, 1095--1153 (2010).

\bibitem{Wang:2026Charmonium}
Z.~Wang, J.~Zhang, and X.~Y.~Wang,
Analysis of $J/\psi$ and $\psi(2S)$ charmonium production
in ultraperipheral lead-lead and proton-lead collisions at LHC energies,
Eur. Phys. J. C \textbf{86}, 668 (2026).

\bibitem{Crede:2024Hyperon}
V.~Crede and J.~Yelton,
70 years of hyperon spectroscopy: A review of strange $\Xi$, $\Omega$ baryons, and the spectrum of charmed and bottom baryons,
Rep. Prog. Phys. \textbf{87}, 106301 (2024).

\bibitem{Hyodo:2021StrangeBaryon}
T.~Hyodo and M.~Niiyama,
QCD and the strange baryon spectrum,
Prog. Part. Nucl. Phys. \textbf{120}, 103868 (2021).

\bibitem{Thiel:2022LightBaryon}
A.~Thiel, F.~Afzal, and Y.~Wunderlich,
Light Baryon Spectroscopy,
Prog. Part. Nucl. Phys. \textbf{125}, 103949 (2022).

\bibitem{wl:2024ext}
X.~Y.~Wang and X.~Liu,
Development Status and Future Plan of Meson Beam Experiments at Home and Abroad,
Nucl. Phys. Rev. \textbf{42}, 10--24 (2025).

\bibitem{Sharov:2011xq}
D.~A.~Sharov, V.~L.~Korotkikh, and D.~E.~Lanskoy,
Phenomenological model for the $\bar{K}N \to K\Xi$ reaction,
Eur. Phys. J. A \textbf{47}, 109 (2011).

\bibitem{Kamano:2014zba}
H.~Kamano, S.~X.~Nakamura, T.~S.~H.~Lee, and T.~Sato,
Dynamical coupled-channels model of $K^-p$ reactions:
Determination of partial-wave amplitudes,
Phys. Rev. C \textbf{90}, 065204 (2014).

\bibitem{Kim:2023DoubleStrangeness}
S.~H.~Kim, J.~K.~Ahn, S.~H.~Kim, S.~I.~Nam, and M.~K.~Cheoun,
Double-strangeness exchange reactions in a hybrid Regge-plus-resonance approach,
Phys. Rev. C \textbf{107}, 065202 (2023).

\bibitem{Luo:2026KXi}
Z.~L.~Luo, J.~J.~Wu, and B.~S.~Zou,
A possible $\Sigma^*$ or $\Lambda^*$ resonance with
$J^P=3/2^-$ in $K^-p\to K\Xi$ scattering,
Phys. Rev. D \textbf{114}, 014062 (2026).

\bibitem{Guo:2025Xi1530}
Q.~Y.~Guo, J.~Liu, P.~Wu, and D.~Y.~Chen,
Production of $\Xi(1530)$ in the $K^-p$ scattering process,
Phys. Rev. C \textbf{112}, 065201 (2025).

\bibitem{Wang:2026KXiMolecule}
K.~Wang, F.~Huang, and B.~S.~Zou,
Exploring $K\Xi^{*}$ and $K^{*}\Xi$ molecular states
and the triangle singularity in the
$K^-p \to K\Xi(1530)$ reaction,
arXiv:2606.29413.

\bibitem{Zhao:2001Kstar}
Q.~Zhao, J.~S.~Al-Khalili, and C.~Bennhold,
Quark model predictions for $K^*$ photoproduction on the proton,
Phys. Rev. C \textbf{64}, 052201(R) (2001).

\bibitem{Oh:2006KstarLambda}
Y.~Oh and H.~Kim,
$K^*$ photoproduction off the nucleon:
$\gamma N\to K^*\Lambda$,
Phys. Rev. C \textbf{73}, 065202 (2006).

\bibitem{Kim:2021HyperonPhoto}
S.~H.~Kim, K.~P.~Khemchandani, A.~Mart\'inez Torres,
S.~I.~Nam, and A.~Hosaka,
Photoproduction of $\Lambda^*$ and $\Sigma^*$ resonances
with $J^P=1/2^-$ off the proton,
Phys. Rev. D \textbf{103}, 114017 (2021).

\bibitem{Kim:2011KstarLambda}
S.~H.~Kim, S.~I.~Nam, Y.~Oh, and H.~C.~Kim,
Contribution of higher nucleon resonances to
$K^*\Lambda$ photoproduction,
Phys. Rev. D \textbf{84}, 114023 (2011).

\bibitem{Wang:2015zcp}
X.~Y.~Wang and J.~He,
$K^{*0}\Lambda$ photoproduction off a neutron,
Phys. Rev. C \textbf{93}, 035202 (2016).

\bibitem{Oh:2006Kstar}
Y.~Oh and H.~Kim,
Scalar $\kappa$ meson in $K^*$ photoproduction,
Phys. Rev. C \textbf{74}, 015208 (2006).

\bibitem{Xiang:2020phx}
J.~Xiang, X.~Y.~Wang, H.~Xu, and J.~He,
Pion-induced $K^*$ production with $\Sigma^*$ baryon
off proton target,
Commun. Theor. Phys. \textbf{72}, 115303 (2020).

\bibitem{Chen:2020Kstar}
S.~F.~Chen and B.~C.~Liu,
$K^*$ production in the $KN\to K\pi p$ reaction,
Chin. Phys. C \textbf{44}, 034107 (2020).

\bibitem{Jackson:2015KXi}
B.~C.~Jackson, Y.~Oh, H.~Haberzettl, and K.~Nakayama,
$\bar{K}N\to K\Xi$ reaction and $S=-1$ hyperon resonances,
Phys. Rev. C \textbf{91}, 065208 (2015).

\bibitem{Anisovich:2017Kstar}
A.~V.~Anisovich \textit{et al.} (CLAS Collaboration),
Differential cross sections and polarization observables from
CLAS $K^*$ photoproduction and the search for new $N^*$ states,
Phys. Lett. B \textbf{771}, 142--150 (2017).

\bibitem{Wang:2026KstarSigmaSDME}
A.~C.~Wang, N.~C.~Wei, and F.~Huang,
Combined analysis of the data on cross sections and spin density
matrix elements for $K^*\Sigma$ photoproduction reactions,
Phys. Rev. C \textbf{114}, 015206 (2026).

\bibitem{Badier:1966KstarXi}
J.~Badier \textit{et al.},
Quelques propri\'et\'es des r\'eactions \`a deux corps
dans les interactions $K^-p$ \`a 3 GeV/$c$,
CEA-R-3037 (1966).

\bibitem{Haque:1966KstarXi}
M.~Haque \textit{et al.}
(Birmingham-Glasgow-London (I.C.)-Oxford-Rutherford Collaboration),
Reactions $K^-p\to {\rm Hyperon}+{\rm Meson}$ at 3.5 GeV/$c$,
Phys. Rev. \textbf{152}, 1148--1161 (1966).

\bibitem{Abrams:1968KstarXi}
G.~S.~Abrams, R.~G.~Glasser, B.~Kehoe, B.~Sechi-Zorn,
and G.~Wolsky,
Final states with two or more strange particles from
4.25-BeV/$c$ $K^-p$ interactions,
Phys. Rev. \textbf{175}, 1697--1704 (1968).

\bibitem{Goldwasser:1970Xi}
E.~L.~Goldwasser and P.~F.~Schultz,
$\Xi^-$ production in 5.5-GeV/$c$ $K^-p$ interactions,
Phys. Rev. D \textbf{1}, 1960--1966 (1970).

\bibitem{Badier:1972Xi}
J.~Badier, E.~Barrelet, G.~R.~Charlton, and I.~Videau,
A search for $\Xi^*$ resonances in $K^-p$ interactions
at 3.95 GeV/$c$,
Nucl. Phys. B \textbf{37}, 429--443 (1972).

\bibitem{Mazzucato:1981Xi}
M.~Mazzucato \textit{et al.}
(Amsterdam-CERN-Nijmegen-Oxford Collaboration),
$\Xi(1320)$ and $\Xi(1530)$ production in $K^-p$
interactions at 4.2 GeV/$c$,
Nucl. Phys. B \textbf{178}, 1--30 (1981).

\bibitem{Bellefon:1975Xi}
A.~de Bellefon, A.~Berthon, P.~Billoir, J.~M.~Brunet,
G.~Tristram, J.~Vrana, B.~Baccari, G.~Poulard,
D.~Revel, and B.~Tallini,
The reactions $K^-p\to\Xi^-K^0\pi^+$ and
$K^-p\to\Xi^-K^+\pi^0$ between 2210 and 2435 MeV
c.m.s. energy,
Nuovo Cim. A \textbf{28}, 289--296 (1975).

\bibitem{Dauber:1969Xi}
P.~M.~Dauber, J.~P.~Berge, J.~R.~Hubbard,
D.~W.~Merrill, and R.~A.~Muller,
Production and decay of cascade hyperons,
Phys. Rev. \textbf{179}, 1262--1285 (1969).

\bibitem{London:1966KstarXi}
G.~W.~London, R.~R.~Rau, N.~P.~Samios, S.~S.~Yamamoto,
M.~Goldberg, S.~Lichtman, M.~Prime, and J.~Leitner,
$K^-p$ interaction at 2.24 BeV/$c$,
Phys. Rev. \textbf{143}, 1034--1091 (1966).

\bibitem{Scheuer:1971Kn}
J.~C.~Scheuer \textit{et al.} (S.A.B.R.E. Collaboration),
Experimental study of two-body and quasi-two-body reactions
in $K^-n$ interactions at 3 GeV/$c$,
Nucl. Phys. B \textbf{33}, 61--83 (1971).

\bibitem{DiBianca:1975XiOmega}
F.~A.~DiBianca and R.~J.~Endorf,
Study of $\Xi^-$ and $\Omega^-$ production from $K^-n$
and $K^-p$ interactions at 4.93 GeV/$c$,
Nucl. Phys. B \textbf{98}, 137--150 (1975).

\bibitem{Nakayama:2006tyg}
K.~Nakayama, Y.~Oh, and H.~Haberzettl,
Photoproduction of $\Xi$ off nucleons,
Phys. Rev. C \textbf{74}, 035205 (2006).

\bibitem{Gao:2010hy}
P.~Gao, J.~J.~Wu, and B.~S.~Zou,
Possible $\Sigma(1/2^-)$ under the $\Sigma^*(1385)$ peak
in $K\Sigma^*$ photoproduction,
Phys. Rev. C \textbf{81}, 055203 (2010).

\bibitem{Wang:2019dsi}
X.~Y.~Wang, J.~He, X.~R.~Chen, Q.~Wang, and X.~Zhu,
Pion-induced production of hidden-charm pentaquarks
$P_c(4312)$, $P_c(4440)$, and $P_c(4457)$,
Phys. Lett. B \textbf{797}, 134862 (2019).

\Needspace{4\baselineskip}
\bibitem{Wang:2024KstarSigma}
X.~Y.~Wang, Y.~Gao, and X.~Liu,
Production potential of hidden-strange molecular pentaquarks
through the $\pi^-p\to K^*\Sigma$ process,
Phys. Rev. D \textbf{111}, 034021 (2025).

\bibitem{Wang:2024BESIII}
X.~Y.~Wang, Y.~Gao, and X.~Liu,
Understanding of the BESIII measurement of
(anti)hyperon-nucleon scattering,
Phys. Lett. B \textbf{862}, 139321 (2025).

\bibitem{Kim:2024HighSpin}
S.~H.~Kim, Y.~Oh, S.~Son, S.~Sakinah, and M.~K.~Cheoun,
Effective Lagrangian for strong and electromagnetic
interactions of high-spin resonances,
Phys. Rev. D \textbf{111}, 054031 (2025).

\bibitem{Pan:2025Lambda}
Y.~Pan and B.~C.~Liu,
Roles of $\Lambda$ resonances in the
$K^-p\to\gamma\Sigma$ reaction,
Phys. Rev. D \textbf{112}, 094053 (2025).

\bibitem{Xie:2014Delta1940}
J.~J.~Xie, E.~Wang, and B.~S.~Zou,
Role of the $\Delta^*(1940)$ in the
$\pi^+p\to K^+\Sigma^+(1385)$ and
$pp\to nK^+\Sigma^+(1385)$ reactions,
Phys. Rev. C \textbf{90}, 025207 (2014).

\bibitem{Schilling:1970}
K.~Schilling, P.~Seyboth, and G.~Wolf,
On the analysis of vector-meson production by polarized photons,
Nucl. Phys. B \textbf{15}, 397--412 (1970).

\bibitem{Jackson:2014KXi}
B.~C.~Jackson, Y.~Oh, H.~Haberzettl, and K.~Nakayama,
Model-independent aspects of the reaction $\bar{K}N\to K\Xi$,
Phys. Rev. C \textbf{89}, 025206 (2014).

\bibitem{Gao:2026Lambda}
Y.~Gao, X.~Y.~Wang, and X.~Liu,
Producing $\Lambda(1405)$ and $\Lambda(1520)$ in
$\pi^-p$ reactions to explore their inner structures,
Phys. Rev. D \textbf{113}, 114006 (2026).

\bibitem{Wang:2024qnk}
X.~Y.~Wang, H.~F.~Zhou, and X.~Liu,
Prospects for detecting the hidden-strange pentaquarklike state
$N^*(2080)$ in the $\pi^-p\rightarrow\phi n$ reaction,
Phys. Rev. D \textbf{110}, 014026 (2024).

\bibitem{Kim:2017nxg}
S.~H.~Kim, S.~I.~Nam, D.~Jido, and H.~C.~Kim,
Photoproduction of $\Lambda^*(1405)$
with the $N^*$ and the $t$-channel
Regge contributions,
Phys. Rev. D \textbf{96}, 014003 (2017).

\bibitem{Bellefon:1977Channels}
A.~de Bellefon, A.~Berthon, P.~Billoir, J.~M.~Brunet,
G.~Tristram, J.~Vrana, B.~Baccari, G.~Poulard,
D.~Revel, and B.~Tallini,
Channel cross-sections of $K^-p$ reactions from
1.934 to 2.516 GeV/$c$,
Nuovo Cim. A \textbf{41}, 451--469 (1977).

\bibitem{Smith:1965Xi}
G.~A.~Smith, J.~S.~Lindsey, J.~Button-Shafer,
and J.~J.~Murray,
Production and decay of $\Xi^*(1820)$,
Phys. Rev. Lett. \textbf{14}, 25--28 (1965).

\bibitem{Merrill:1970Xi}
D.~Merrill \textit{et al.} (S.A.B.R.E. Collaboration),
Cross sections of final states produced in $K^-N$
interactions at 3 GeV/$c$,
Nucl. Phys. B \textbf{18}, 403--424 (1970).

\bibitem{Chen:2026HHaS}
X.~R.~Chen \textit{et al.},
Huizhou Hadron Spectrometer: A Proposed High-Rate Experimental
Setup at the High-Intensity Heavy-Ion Accelerator Facility,
Nucl. Sci. Tech. \textbf{37}, 156 (2026).

\bibitem{Aoki:2021cqa}
K.~Aoki, H.~Fujioka, T.~Gogami, Y.~Hidaka, E.~Hiyama,
R.~Honda, A.~Hosaka, Y.~Ichikawa, M.~Ieiri, M.~Isaka,
\textit{et al.},
Extension of the J-PARC Hadron Experimental Facility:
Third White Paper,
arXiv:2110.04462.

\bibitem{Adams:2018pwt}
B.~Adams, C.~A.~Aidala, R.~Akhunzyanov, G.~D.~Alexeev,
M.~G.~Alexeev, A.~Amoroso, V.~Andrieux, N.~V.~Anfimov,
V.~Anosov, A.~Antoshkin, \textit{et al.},
Letter of Intent: A New QCD Facility at the M2 Beam Line
of the CERN SPS (COMPASS++/AMBER),
arXiv:1808.00848.

\bibitem{Dobbs:2022KLF}
S.~Dobbs (KLF),
Strange Hadron Spectroscopy with the KLong Facility at
Jefferson Lab,
Rev. Mex. Fis. Suppl. \textbf{3}, 0308032 (2022).




\end{thebibliography}
\end{document}